\documentclass[%
 aip,
 amsmath,amssymb,
 reprint,%
]{revtex4-1}

\usepackage{graphicx}
\usepackage{dcolumn}
\usepackage{bm}

\usepackage[utf8]{inputenc}
\usepackage[T1]{fontenc}
\usepackage{mathptmx}
\usepackage{etoolbox}
\usepackage{diagbox}
\usepackage{url}
\usepackage{color}

\makeatletter
\def\@email#1#2{%
 \endgroup
 \patchcmd{\titleblock@produce}
  {\frontmatter@RRAPformat}
  {\frontmatter@RRAPformat{\produce@RRAP{*#1\href{mailto:#2}{#2}}}\frontmatter@RRAPformat}
  {}{}
}%

\newcommand{\tonde}[1]{\left(#1\right)}
\newcommand{\quadre}[1]{\left[#1\right]}

\newcommand{\graffe}[1]{\left\{#1\right\}}

\newcommand{\be}{\begin{equation}}
\newcommand{\ee}{\end{equation}}
\newcommand{\ba}{\begin{eqnarray}}
\newcommand{\ea}{\end{eqnarray}}
\newcommand{\y}{{\mathbf{y}}}
\newcommand{\Y}{{\mathbf{Y}}}
\newcommand{\Yt}{{\mathbf{Y}^{\tonde{t}}}}

\newcommand{\Yijt}{{Y_{ij}^{\tonde{t}}}}
 \newcommand{\Yij}{{Y_{ij}}}

\newcommand{\etm}{^{\tonde{t-1}}}
\newcommand{\et}{^{\tonde{t}}}

\newcommand{\etprime}{^{\tonde{t}\prime}}
\newcommand{\etp}{^{\tonde{t+1}}}

\newcommand{\binpar}{\theta}

\newcommand{\ibinpar}{{\overleftarrow{\binpar}}}
\newcommand{\obinpar}{{\overrightarrow{\binpar}}}

\newcommand{\argmax}[1]{\underset{#1}{\operatorname{arg}\,\operatorname{max}}\;}

\makeatother
\begin{document}

\preprint{AIP/123-QED}

\title[Score-Driven Exponential Random Graphs]{Score-Driven Exponential Random Graphs:\\ A New Class of Time-Varying Parameter Models for Temporal Networks}
\author{D. Di Gangi}
\affiliation{Domotz, via U. Forti 1, 56121 Pisa, Italy}%
\author{G. Bormetti}%
\affiliation{ 
Department of Economics and Management, University of Pavia, Via San Felice al Monastero 5, 27100, Pavia, Italy
}%
\email{giacomo.bormetti@unipv.it}
\author{F. Lillo}
\affiliation{%
Department of Mathematics, University of Bologna, Piazza di Porta San Donato 5, 40126, Bologna, Italy, and Scuola Normale Superiore, Piazza dei Cavalieri 7, 56126, Pisa, Italy
}%

\date{\today}

\begin{abstract}
Motivated by the increasing abundance of data describing real-world networks that exhibit dynamical features, we propose an extension of the Exponential Random Graph Models (ERGMs) that accommodates the time variation of its parameters. 
Inspired by the fast-growing literature on Dynamic Conditional Score models, each parameter evolves according to an updating rule driven by the score of the ERGM distribution. We demonstrate the flexibility of score-driven ERGMs (SD-ERGMs) as data-generating processes and filters and show the advantages of the dynamic version over the static one. 
We discuss two applications to temporal networks from financial and political systems. First, we consider the prediction of future links in the Italian interbank credit network. Second, we show that the SD-ERGM allows discriminating between static or time-varying parameters when used to model the U.S. Congress co-voting network dynamics. 
\end{abstract}

\maketitle

\begin{quotation}
The paper introduces an innovative, dynamic model for temporal networks. The novel methodology is strongly interdisciplinary, and the estimation of model parameters is straightforward. We present two examples from financial and political systems, supporting the approach's flexibility and showcasing its potential widespread applicability.
\end{quotation}

\section{\label{sec:introduction}Introduction}

A network is a useful abstraction for a system composed of several single elements with some pairwise relations. The simplified description of social, economic, biological, and transportation systems, often very complex, in terms of nodes and links, attracted and still attracts an enormous amount of attention~\citep{RevModPhys.74.47,bullmore2009complex,newman2010networks,easley2010networks,allen2009networks}.  
Formally, a network $G$ is a pair $ \left( V, E \right) $ where $V$ is a set of nodes and  $E$ is a set of node pairs named links. The nodes are labeled, and a link is identified by the pair of nodes it connects $(i,j)$. To each $G$, we can assign an adjacency matrix $\Y$ such that $\Yij = 1 $ if link $\tonde{i,j}$ is present in $E$ and $\Yij = 0 $ otherwise. 
Links may have orientations. The corresponding network is dubbed directed. If the elements of the adjacency matrix are allowed to be different from $0$ or $1$, one speaks of weighted networks. In the following, we will focus on directed networks rather than consider the weighted variant.  

Often, systems that are fruitfully described as networks evolve in time. When the number of nodes and/or pairwise interactions change over time, one usually speaks of temporal networks~\citep{holme2012temporal,craig2014interbank,rossetti2018community}. This paper will focus on temporal networks where links evolve in discrete time. A temporal network is a sequence of networks, each associated with an adjacency matrix and observed at $T$ different points in time. The whole time series is given in terms of a sequence of matrices $\graffe{\Yijt}_{t=1}^{T}$. 

We introduce an innovative approach to temporal networks based on two main ingredients: (i) a parametric probabilistic model, according to which one can sample a network realization. A natural choice is the class of statistical models for networks known as Exponential Random Graph Models (ERGMs). (ii) A simple mechanism to induce dynamics on the network sequence by introducing time variation on the model parameters. The Dynamic Conditional Score (DCS) approach provides a flexible candidate. Our extension of the ERGM framework allows model parameters to change over time in a score-driven fashion. We develop a new class of temporal network models and show its versatility and effectiveness in capturing time-varying features. The information encoded in $\mathcal{F}_{t-1}$ is exploited to filter the time-varying parameters (tvps) $\theta\et$ at time $t$. We refer to this class as \textit{Score-Driven Exponential Random Graph Models} (SD-ERGMs). A generic SD-ERGM can be used either as a data-generating process (DGP) to sample synthetic sequences of graphs or as an effective filter of latent tvps, regardless of the true DGP. 

We are by no means the first to discuss models for temporal networks. For a review of latent space temporal network models, one can refer to Ref.~\onlinecite{kim2018review} Extensions of the ERGM framework for the description of temporal networks exist in the literature. Two are the main streams. The first one, termed TERGM, was pioneered in Ref.~\onlinecite{robins2001random} and further explored in Refs.~\onlinecite{hanneke2010discrete,cranmer2011inferential,krivitsky2014separable}. This approach builds on the ERGM but allows the network statistics to define the probability at time $t$ to depend on current and previous networks up to time $t-K$. This $K$-step Markov assumption is a defining feature of the TERGMs. 
A second approach allows for the parameters of the ERGM to be time-varying. A notable example is the Varying-Coefficient-ERGM~\cite{lee2020varying}, allowing for smooth parameter time variation. The approach differs from ours in several respects. Specifically, to infer parameter time-variation at time $t$, it uses all the available observations, including those from future times $t^\prime > t$~\footnote{In time series jargon, it is a smoother and not a filter.}. Consequently, it cannot be used to draw causal sequences of time-evolving networks. A related but different approach  \cite{mazzarisi2020dynamic} considers the possibility of a random evolution of node-specific parameters. As a crucial difference with the SD-ERGM, the parameter evolution is driven by an exogenous source of randomness. Following the language of Ref.~\onlinecite{cox1981statistical}, the approach is \textit{parameter driven}, while we consider an \textit{observation driven} dynamics. Finally, it is important to mention that the social science literature has considered alternative frameworks for modeling temporal networks. Notable examples are the Stochastic Actor Oriented model~\cite{snijders1996stochastic} and the Relational Event Model~\cite{butts20084}. For an overview of contributions, we refer to the literature therein.

The rest of the paper is organized as follows. In Section \ref{sec:intro_concepts}, we review some key concepts on ERGMs and observation-driven models. In Section~\ref{sec:methodology_SDERGM}, we introduce the new class of models and validate it with extensive numerical experiments for three specific instances of the SD-ERGM. Section~\ref{sec:applications} presents the results from an application to two real temporal networks: The e-MID interbank network for liquidity supply and demand and the U.S. Congress co-voting political network. Section~\ref{sec:conclusions} draws the relevant findings.   

\section{Main building blocks} \label{sec:intro_concepts}
The two main ingredients in our approach are the ERGMs, a static class of network models well-known in physics, and the DCSs, a recent development in time-series econometrics.
   
\subsection{ERGM - Exponential Random Graph Models}\label{sec:lit_rev_ergm}
A statistical network model can be specified by providing the probability distribution over the set of possible adjacency matrices~\cite{Kolaczyk:2009:SAN:1593430}. If the distribution belongs to the exponential family~\citep{barndorff2014information}, then the model is named ERGM, and its log-likelihood takes the form 
\begin{equation}
\label{eq:ergm_def}
\log{P\tonde{\Y}} = \sum_{s} \theta_s h_s\tonde{\Y}  - \log{\tonde{\mathcal{K}\tonde{\theta}}}, 
\end{equation}
where $h$ are network statistics, $\theta$ is the vector of parameters whose component $\theta_s$ is associated with the network statistic $h_s\tonde{\Y}$, and $\mathcal{K}\tonde{\theta} = \sum_{\graffe{\Y}} e^{   \theta_s h_s\tonde{\Y} } .$
The ERGMs literature is vast and still growing \citep{schweinbergerexponential}. 
The ERGM framework is intrinsically linked to the very well-known \textit{principle of maximum entropy}  \citep{Shannon:2001:MTC:584091.584093} and its applications to statistical physics \citep{PhysRev.106.620}. Indeed, an ERGM with sufficient statistics $h\tonde{\theta}$ naturally arises when looking for the probability distribution which maximizes the entropy under a linear equality constraint on the statistics $h\tonde{\theta}$~\citep{PhysRevE.70.066117,PhysRevE.78.015101}.  
The sufficient statistics $h_s\tonde{\Y}$, known as network statistics, are functions of the adjacency matrix $\Y$, whose entries are binary random variables. The probability mass function (PMF) is defined by \eqref{eq:ergm_def}.
The normalizing factor $\mathcal{K}\tonde{\theta }$ is often unavailable as a closed-form function of the parameters $\theta $. 

In the following, we will focus on two specific examples of ERGMs that describe distinct features of the network and require different approaches to parameter inference. The first one is meant to capture the heterogeneity in the number of connections each node can have, and it allows for straightforward maximum likelihood estimation (MLE)~\citep{chatterjee2011random}. It is known as \textit{beta model}, \textit{fitness model}, and \textit{configuration model} \citep[][]{zermelo1929berechnung,Holland81anexponential,PhysRevLett.89.258702,PhysRevE.70.066117,PhysRevE.78.015101,chatterjee2011random}. The second one is an ERGM having as statistic the Geometrically Weighted Edgewise Shared Partners (GWESP).
That is a network statistic describing transitivity in the formation of links, i.e., the tendency of connected nodes to have common neighbors and belongs to a family of network statistics referred to as curved exponential random graphs, proposed in Refs.~\onlinecite{snijders2006new,robins2007recent} and discussed in Ref.~\onlinecite{hunter2006inference}. In the latter case, the inference is complicated because the normalizing factor in \eqref{eq:ergm_def} is not available in closed form. In such cases, there are two standard approaches to ERGM inference, both consisting of maximizing alternative functions that are known to share the same optimum as the exact likelihood. In Appendix~\ref{sec:appendix_ERGMs}, we provide more details on the beta model and GWESP ERGMs definitions as long as more details on the associated inference procedures. 

\subsection{Score-Driven Models}\label{sec:lit_rev_score_driven}
The second main ingredient of this work is the class of DCS models~\cite{creal2013generalized,harvey2013dynamic}, also known as Generalized Autoregressive Score models~\footnote{\texttt{http://www.gasmodel.com/index.htm} for the updated collection of papers dealing with GAS models.}. 
In the language of Ref.~\onlinecite{cox1981statistical}, DCSs belong to observation-driven models. Let us consider a sequence of observations $\graffe{y\et}_{t=1}^T$, where each $y\et \in\mathbb{R}^M$,  and a conditional probability density $P\tonde{y\et\vert f\et}$, that depends on a vector of tvps $f\et \in \mathbb{R}^K$. Defining the score as 
\begin{equation}\label{eq:score_def}
\nabla\et  = \frac{\partial \log{P\tonde{y\et\vert f\et}}}{\partial f\etprime}\,,
\end{equation}
a Score-Driven model assumes that the recursive relation
\be  \label{eq:gasupdaterule}
f\etp = w +\boldsymbol{\beta} f\et + \boldsymbol{\alpha} \boldsymbol{S}\et \nabla\et , \\
\ee 
rules the time evolution of $f\et$, with $w$, $\boldsymbol{\alpha}$ and $\boldsymbol{\beta}$ are static parameters, $w$ being a $K$ dimensional vector and $\boldsymbol{\alpha}$ and $\boldsymbol{\beta}$  $K\times K$ matrices.  $\boldsymbol{S\et}$ is a $K\times K$ scaling matrix usually chosen as a power of the inverse of the Fisher information matrix associated with $P\tonde{y\et\vert f\et}$.

The parameter updating rule can be intuitively motivated by general assumptions based on information theory principles. One can assume that the network behavior varies based on surprise: The more an observation of the network’s state, i.e., the adjacency matrix, is ``unexpected", the more the relations between its components will change. The most common
measure of surprise is minus the logarithm of the likelihood of observing the current state conditional on the level of the model parameters. As a second principle, one assumes that the reaction to surprise is to adapt to it, making what had been unexpected at that moment less surprising in the future. This implies that the parameters change to minimize the surprise, i.e., increase the log-likelihood of the last observation and thus move along the steepest direction the gradient provides. Then, the updated parameter value will be a linear combination of the current value and the log-likelihood score.

The structure of the conditional observation density determines the score, from which the dependence of $f\etp$ on the vector of observations $y\et$ follows. When the model is viewed as a DGP, the update results in stochastic dynamics precisely thanks to the random occurrence of $y\et$. When the score-driven recursion is regarded as a filter, the update rule in \eqref{eq:gasupdaterule} is used to obtain a sequence of filtered  $\graffe{\hat{f}\et}_{t=1}^T$. In this setting, one estimates the static parameters by maximizing the log-likelihood of the whole sequence of observations.

A second look at eq. \eqref{eq:gasupdaterule} reveals the similarity of the score-driven recursion with the iterative step from a Newton algorithm, whose objective function is precisely the log-likelihood function. As mentioned above, at each step, the score pushes the parameter vector along the log-likelihood steepest direction. Moreover, there are motivations, grounded on the variation of the Kullback-Leibler divergence, for the optimality of the score-driven updating rule \cite{blasques2015information}, as we review in Appendix~\ref{sec:appendix_DCS}.

Many well-known econometrics models can be expressed as Score-Driven models. Famous examples are the Generalized Autoregressive Conditional Heteroskedasticity (GARCH) model~\cite{BOLLERSLEV1986307}, the Exponential GARCH model~\cite{nelson1991conditional}, the Autoregressive Conditional Duration model~\cite{engle1998autoregressive}, and the Multiplicative Error Model~\cite{engle2002new}. The introduction of this framework in its full generality opened the way to applications in various contexts. 

Before moving to the most crucial section, let us mention two relevant technical aspects for the applications discussed in the paper: the computation of the confidence bands of the filtered parameters and testing for the parameter temporal variation. We postpone the technical discussion to Appendix~\ref{sec:appendix_DCS} for brevity.

\section{Score-Driven Exponential Random Graphs }\label{sec:methodology_SDERGM}

This section describes the methodological innovation introduced by the manuscript. We present the general SD-ERGM framework, discuss the score-driven approach's applicability to three different ERGMs in detail, and validate their performances with extensive numerical simulations.


We apply the score-driven methodology to ERGMs to allow any of the parameters $\theta_s$ in \eqref{eq:ergm_def} to have a stochastic evolution driven by the score of the static ERGM model, computed at different points in time. This approach results in a framework for describing temporal networks, more than in a single model, in the same way ERGM is considered a modeling framework for static networks.

Conceptually, applying the score-driven approach is pretty straightforward. Given the observations $\graffe{\Yijt}_{t=1}^{T}$,  we can apply the update rule in \eqref{eq:gasupdaterule} to all or some elements of $\theta$, each of which is associated with a network statistic in \eqref{eq:ergm_def}. 
To do this, we need to compute the derivative of the log-likelihood at every time step, i.e., for each adjacency matrix $\Y\et$.  For the general ERGM, the elements of the score take the form 
\begin{equation*}
\nabla_s\et\tonde{\theta}  =  h_s\tonde{\Yt}  - \frac{\partial \log{\mathcal{K}\tonde{\theta}} }{\partial \theta_s}\,,
\end{equation*}
and  the vector of tvps evolves according to~\eqref{eq:gasupdaterule} with $f\et$ replaced by $\theta\et$.
Hence, conditionally on the value of the parameters $\theta\et$ at time $t$ and the observed adjacency matrix $\Y\et$, the parameters at time $t+1$ are deterministic. When used as a DGP, the SD-ERGM describes stochastic dynamics because, at each time $t$, the adjacency matrix is not known in advance but must be randomly sampled from $P\tonde{\Yt\vert \theta\et}$ and used to compute the score.
When the sequence of networks $\graffe{\Yt}_{t=1}^T$ is observed, the static parameters  $\tonde{w,\boldsymbol{\beta},\boldsymbol{\alpha}}$, that best fit the data, can be computed via MLE. Taking into account that each network $\Yt$ is independent of all the others \textit{conditionally} on the value of $\theta\et$, the log-likelihood can be written as
\begin{eqnarray} \label{eq:SDERGM_likelihood}
&\log{P}\tonde{\graffe{\Yt}_{t=1}^T \vert w,\beta,\alpha} = \nonumber\\
&\sum_{t=1}^T \log{P\tonde{ \Yt  \vert \theta\et\tonde{w,\beta,\alpha,\graffe{\Y^{\tonde{t^\prime}}}_{t^\prime=1}^{t-1}} }}.
\end{eqnarray}
The computation of the normalizing factor and its derivative with respect to the parameters is essential for the SD-ERGM. Not only does it enter the definition of the update, but it is also required to optimize~\eqref{eq:SDERGM_likelihood}. 

Our primary motivation for introducing the SD-ERGM is to describe the time evolution of a sequence of networks using the evolution of the parameters of an ERGM. From the context or previous studies of static networks in terms of ERGM, we assume we know which statistics are more appropriate in describing a given network. Hence, we do not discuss the choice of statistics in the context of temporal networks but refer the reader to Refs.~\onlinecite{goodreau2007advances,hunter2008goodness,SHORE201516} for examples of feature selection and goodness-of-fit evaluation.

In this final paragraph, we anticipate the SD extensions of ERGMs with given statistics detailed in the following pages. The first example allows for the exact computation of the likelihood, but the number of parameters can become significant for a large network. The second example discusses how an SD-ERGM can be defined when the log-likelihood is not known in closed form.  Using extensive numerical simulations, we show that SD-ERGMs are very efficient at recovering the paths of tvps when the DGP is known, and the score-driven model is employed as a misspecified filter. Moreover, we show the first application of the Lagrange Multiplier (LM) test~\citep{calvori2017testing} in assessing the time-variation of ERGM parameters. 

\subsection{Score-Driven Beta Model } \label{sec:beta_model}

Our first specific example is the Score-Driven version of the \textit{beta model}, introduced in Sec.~\ref{sec:lit_rev_ergm} and further discussed in Appendix~\ref{sec:appendix_ERGMs}. We start with this model because of its wide applications and relevance in various streams of literature and because the likelihood of the ERGM and its score can be computed exactly. Moreover, the number of local statistics, the degrees, and parameters can become very large for large networks. Since we must describe the dynamics of many parameters, this last feature challenges any time-varying parameter version of the beta model. At the end of this Section, we will show how the SD framework allows for a parsimonious description of such high-dimensional dynamics. 

As anticipated, the SD-beta model is defined by applying \eqref{eq:gasupdaterule} to each of the $\obinpar$ and $\ibinpar$ parameters. Among the possible choices, we use as scaling the diagonal matrix  $S_{ij}\et = {\delta_{ij} I_{ij}\et}^{-1/2} $, where $\boldsymbol{I\et} = {\mathbb E}[{\nabla\et {\nabla\et}^\prime}] $, i.e., we scale each element of the score by the square root of its variance. It is widespread, in score-driven models with numerous tvps, to restrict the matrices $\boldsymbol{\alpha}$ and $\boldsymbol{\beta}$ of~\eqref{eq:gasupdaterule} to be diagonal. In this work, we consider a version of the score update having only three static parameters $\tonde{w_s,\beta_s,\alpha_s}$ for each dynamical parameter $\theta_s$.  
The resulting update rule for the beta model is  
\begin{eqnarray}\label{eq:sd_beta_update}
	\ibinpar_s\etp &= {w}_s^{\text{in}} + {\beta}^{\text{in}}_s 	\ibinpar_s\et +  {\alpha}^{\text{in}}_s \frac{\sum_{i} \tonde{Y_{is}\et - p_{is}\et} }{\sqrt{\sum_i p_{is}\et\tonde{1 - p_{is}\et }}} \nonumber\\
	\obinpar_s\etp &= {w}_s^{\text{out}} + {\beta}^{\text{out}}_s 	\obinpar_s\et +  {\alpha}^{\text{out}}_s \frac{\sum_{i} \tonde{Y_{si}\et - p_{si}\et} }{\sqrt{\sum_i p_{si}\et\tonde{1 - p_{si}\et }}},
\end{eqnarray}
where the superscripts $in$ and $out$ indicate the first and second half of the parameter vectors, respectively.
To simplify the inference procedure, we consider a two-step approach. First, we fix the node-specific parameters $w_i$ to target the unconditional means of $\ibinpar$ and $ \obinpar $ resulting from an ERGM with static parameters. Conditionally on the target values, we estimate the remaining parameters $\alpha^\text{in}$, $\alpha^\text{out}$, $\beta^\text{in}$, and $\beta^\text{out}$. We verified that the bias introduced by the two-step procedure is negligible, and results remain similar when the joint estimation is performed.

\subsubsection{SD-ERGMs as filters: Numerical Simulations} \label{sec:sim_sd_beta}
As mentioned in the Introduction, SD-ERGMs (as other observation-driven models, e.g., GARCH) can be seen as DGPs or predictive filters~\cite{Nelson96} since tvps follow one-step-ahead predictable processes. In this Section, we show the ability of the ERGMs in the latter setting. Specifically, we simulate generic non-stationary evolution for temporal network parameters $\theta\et$.
We then use the SD-ERGM to filter the paths of the parameters and evaluate its performances. It is important to note that the parameters' simulated dynamics differ from the score-driven ones specified for the estimation.

In practice, at each time $t$, we sample the adjacency matrix from the PMF of an ERGM with parameters\footnote{In the following, the notation with a bar refers to the true parameters used in the DGP.}  $ \bar{\theta}\et $, evolving according to known temporal patterns that define different DGPs. We then use the realizations of the sampled adjacency matrices to filter the patterns. We consider a sequence of $T = 250$ time steps for a network of $10$ nodes, each with parameters $\overline{\ibinpar_i}\et $ and $\overline{\obinpar_i}\et $ evolving with predetermined patterns. We test four different DGPs. The first one is a naive case with constant parameters $\overline{\theta}\et = \overline{\theta}_{0}$. The elements of $\overline{\theta}_{0}$ are chosen to ensure heterogeneity in the expected degrees of the nodes under the static beta model. For the remaining three DGPs, half of the parameters are static, and half are time-varying, evolving with either a deterministic sinusoidal function, a deterministic step function, or a stochastic AR(1) dynamics. More details on the definition of such DGPs are given in the Appendix~\ref{sec:appendix_num_details}. 

In the following, we benchmark the performance of the SD-ERGMs with that of a sequence of cross-sectional estimates of static ERGMs, i.e., one ERGM estimated for each $t$. We quantify the performance of the two approaches computing the  Root Mean Square Error $ \frac{1}{T} \sqrt{\sum_t \tonde{\bar{\theta}_s\et - \hat{\theta}_s\et  }^2 }$, that describes the distance between the known simulated path and the filtered.
We then average the RMSE across all the tvps and $100$ simulations and report the results in Table \ref{tab:beta_mod_RMSE}. These results confirm that the SD beta model outperforms the standard beta model in recovering the true time-varying pattern.  Notably, this holds even when the DGP is inherently nonstationary, as in the case of the DGP, where each parameter has a step-like evolution. Indeed, the results of this Section and Section~\ref{sec:SDERGM_glob_Stat} confirm that, while the SD update rule~\eqref{eq:gasupdaterule} defines a stationary DGP \citep{creal2013generalized}, using SD models as filters, we can effectively recover nonstationary parameters' dynamics.
\begin{table}  
\caption{\label{tab:beta_mod_RMSE} RMSEs (on a percentage base) of the filtered paths averaged over all tvps and all Monte Carlo replicas of the numerical experiment. Left column: results from the cross-sectional estimates of the beta model; right column: score-driven beta model results. Each row corresponds to one of the four DGPs.} 
\centering
\begin{tabular}{l|cc}
	 DGPs & \multicolumn{2}{c}{\rule{0pt}{15pt} Average RMSE  }   \\ 
	\rule{0pt}{15pt}  &   beta model   &   SD-beta model     \\
	\hline
	\rule{0pt}{12pt}  Const &  1.75  & 0.20  \\
	 
	\hline
	\rule{0pt}{12pt} Sin & 2.76  & 0.34 \\
 
	\hline
	\rule{0pt}{12pt}  Steps &2.46  & 0.28   \\
 
	\hline
	\rule{0pt}{12pt}  AR(1)  &1.82  & 0.24 \\
 
\end{tabular}
\end{table} 
Our last numerical simulations for the SD beta model explore its applicability and performance for networks of increasing size. 
We explore this setting for two reasons. The first one is that networks with a large number of nodes describe many real systems. The second reason is that we want to compare the performance of our approach with that of the standard beta model in regimes where the latter is known to perform better under suitable conditions. Indeed, as mentioned in Appendix~\ref{sec:appendix_ERGMs}, asymptotic results on the single observation estimates \cite{chatterjee2011random} guarantee that, if the network density remains constant as $N$ grows, the accuracy of the cross-sectional estimates increases. We want to check numerically that, within the regime of dense networks, the accuracy of the static and SD versions of the beta model reaches the same level. To check whether the SD approach provides any advantage for large networks, we perform numerical experiments similar to the previous ones but in a different and more realistic regime of sparse networks, i.e., keeping the average degree constant. Moreover, to ease the computational burden for the estimates, we consider a restricted version of the SD-Beta model, as detailed in Appendix~\ref{sec:appendix_SDBeta}, having only one set of parameters $\tonde{{\beta}^{\text{in}}, {\beta}^{\text{out}}, {\alpha}^{\text{in}}, {\alpha}^{\text{out}}}$ for the whole network, instead of one set per each node. 

This analysis considers only one dynamical DGP and many different values of $N$. Among the DGPs used above, we focus on the one with smooth and periodic time variation. Most importantly, we set a maximum degree attainable for a node and let it depend on $N$ in two distinct ways, each corresponding to a different density regime: one generating \textit{sparse} networks and the other \textit{dense} ones. It is worth noticing that the asymptotic results of Ref.~\onlinecite{chatterjee2011random} are expected to hold only in the dense case.
\begin{figure}
	\includegraphics[width=0.45\linewidth 	]{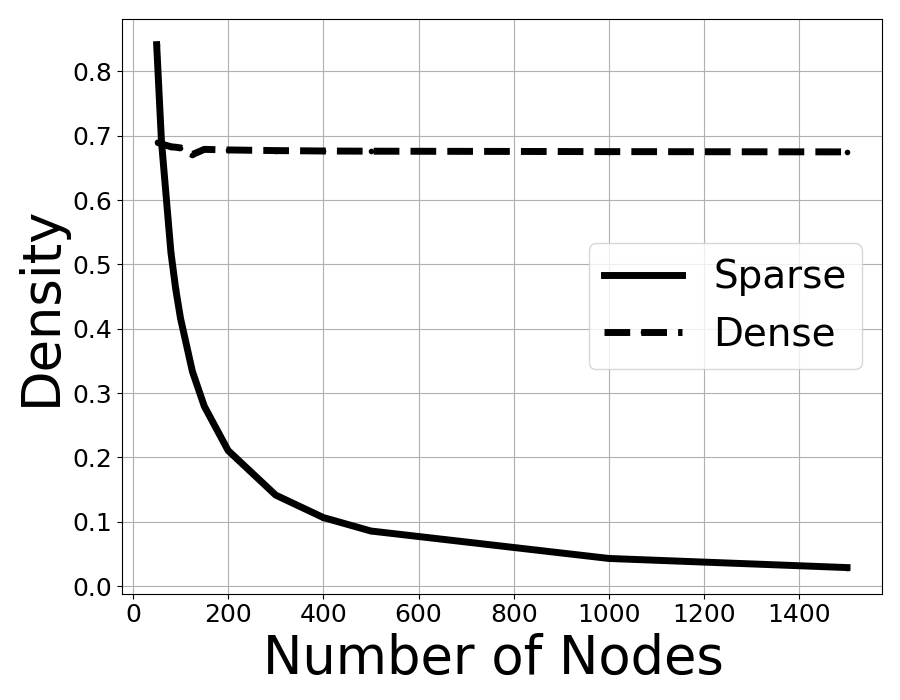} 
	\includegraphics[width=0.45\linewidth 	]{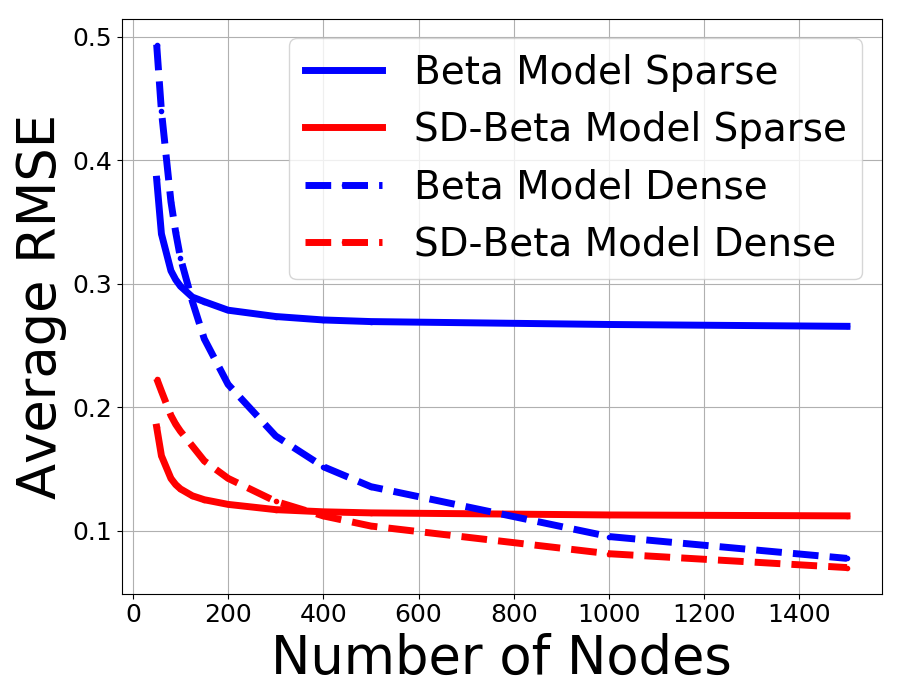}
	\caption{Left panel: average density as a function of the number of nodes $N$ in the dense (dashed line) and sparse (solid line) regimes. Right panel: average RMSE of the filtered parameters with respect to the simulated DGP in both the dense (dashed lines) and sparse (solid lines) regimes. The average RMSE from the ERGM is plotted in blue, while the one from the SD-ERGM is in red.}\label{fig:rmse_beta_model_large_N} 
\end{figure}
The average densities for different values of $N$ in the two regimes are shown in the left panel of Figure \ref{fig:rmse_beta_model_large_N}. Then, for both regimes and each value of $N$, we compute the average RMSE across all tvps and all Monte Carlo replicas. In the right panel of Figure~\ref{fig:rmse_beta_model_large_N}, the average RMSEs for different values of $N$ indicate that, also for large networks, the SD version of the beta model attains better results compared with the cross-sectional estimates. As expected, both approaches reach the same accuracy in the dense network regimes as long as $N$ becomes larger. However, in the more realistic sparse regime, the performance of the SD-ERGM remains much superior for both small and large network dimensions.

\subsection{Pseudo-Likelihood SD-ERGM}\label{sec:SDERGM_glob_Stat}

As mentioned earlier, the dependence of the normalizing function on the $\theta$ parameters is often unknown. This fact prevents us from computing the score function and directly applying the update rule~\eqref{eq:gasupdaterule} to a large class of ERGMs. To circumvent this obstacle, instead of the unattainable score of the exact likelihood, we propose to use the score of the pseudo-likelihood, discussed in Sec. \ref{sec:lit_rev_ergm}, that we refer to as pseudo-score 
\be\label{eq:pseudoscore}
\frac{\partial \log{PL\tonde{\Yt\vert \theta}}}{\partial \theta_s\etprime}  = \sum_{ij}  \delta_{ij}^s\tonde{\Yijt  - \frac{1 }{1 + e^{ - \sum_l \theta_l \delta_{ij}^l}} } ,
\ee
in place of the exact score in the definition of the SD-ERGM update~\eqref{eq:gasupdaterule}. 
Additionally, we use the pseudo-likelihood for each observation $\Y\et$ in \eqref{eq:SDERGM_likelihood} to infer the static parameters.

Our approach, based on the score of the pseudo-likelihood, requires as input the change statistics for each function $h_s\tonde{\Yt}$~\footnote{For practical applications, it is very convenient that, for a large number of network functions, an efficient implementation to compute change statistics is made available in the R package \textit{ergm}~\cite{hunter2008ergm}.}.  
In the following, we show that the update based on the pseudo-likelihood score effectively filters the path of tvps. Remarkably, this is true even when the probability distribution in the DGP is exact, i.e., when we sample from the exact likelihood and then use the SD-ERGM based on the pseudo-likelihood to filter.

\subsubsection{SD-ERGM for Transitivity and Network Density }\label{sec:GWESP_numerical_simul}
In this section, we discuss numerical simulations for an ERGM whose normalization is not known in closed form, which we also apply to real data in Section \ref{sec:us_congr_app}. 
We show the concrete applicability of the SD-ERGM approach based on the pseudo-score and its performance as a filter compared with the cross-sectional MCMC estimates of the standard ERGM. The models we consider have two statistics. The first one is the total number of links present in the network. The second statistic is the GWESP, introduced in Section \ref{sec:lit_rev_ergm}. The ERGM is thus defined by
\be\label{eq:ERGM_glob_def}
\sum_{s} \theta_s h_s\tonde{\Yt} = \theta_1 \sum_{ij} \Yijt + \theta_2 \text{GWESP}\tonde{\Yt}\,.
\ee
To test the efficiency of the SD-ERGM, we simulate a known temporal evolution for the parameters and, at each time step, we sample the exact PMF from the resulting ERGMs.  Finally, we use the observed change statistics for each time step to estimate two alternative models: a sequence of cross-sectional ERGMs and the SD-ERGM. In what follows, we indicate the values from the DGP of parameter $s$ at time $t$ as $\bar{\theta}_s\et$.

\begin{figure}[t!]
    \includegraphics[width=\linewidth]{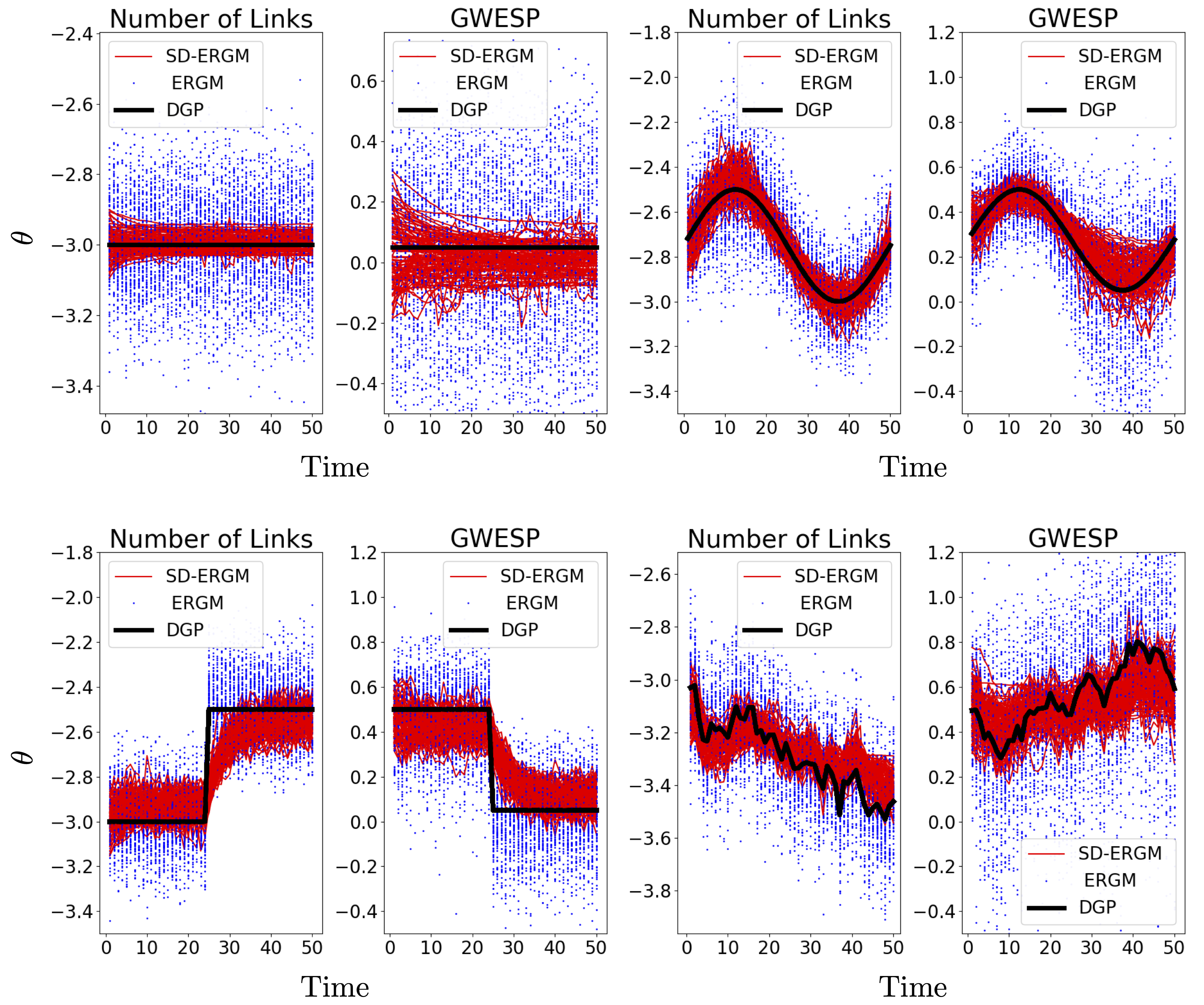}
	\caption{Filtered paths of the parameters of the ERGM in \eqref{eq:ERGM_glob_def} with tvps. The path from the true DGP is in black. The blue dots are the cross-sectional ERGM estimates, and the red lines are the SD-ERGM filtered paths.}\label{fig:global_stats_filter_paths} 
\end{figure}

We investigate four DGPs similar to those analyzed in Section \ref{sec:beta_model}. We sample and estimate the models 50 times for each DGP. Figure~\ref{fig:global_stats_filter_paths} compares the cross-sectional estimates and the score-driven filtered paths.
Table \ref{tab:mse_filter_gwesp} reports the RMSE of the GWESP tvps, averaged over the different realizations for the whole sequence $t=1,2,\dots,T$. The SD-ERGM  outperforms the cross-sectional ERGM estimates for all the investigated time-varying patterns. Moreover, when the constant DGP is considered, i.e., $\bar{\theta}_1\et = \bar{\theta_1} $ and   $\bar{\theta}_2\et = \bar{\theta_2} $, the average RMSE of the SD-ERGM is larger but comparable, than the correctly specified ERGM that uses all the longitudinal observations to estimate the parameters. The latter result confirms that the SD-ERGM is a reliable and consistent choice even for the static case.
\begin{table} 
\caption{\label{tab:mse_filter_gwesp}  First four columns: RMSEs for the filtered paths of the tvps, averaged over 50 repetitions, for the evolutions of Figure \ref{fig:global_stats_filter_paths}. The last three columns describe the accuracy of the test for dynamics in the parameters, considering the DGPs in Figure \ref{fig:global_stats_filter_paths}, as well as alternative DGPs where only one parameter is time-varying. We report the percentage of times that the LM test correctly identifies the parameter as time-varying (or static in the case of the first DGP). The chosen threshold for the p-values is $0.05$.} 
\centering
\begin{tabular}{l||cc|cc||ccc}
\rule{0pt}{20pt}  DGP &  \multicolumn{4}{c}{\rule{0pt}{15pt} Average RMSE }  & \multicolumn{2}{c}{\rule{0pt}{15pt}LM Test }  \\ 
\hline  \\ 
&   \multicolumn{2}{c}{ERGM}   &  \multicolumn{2}{c}{ SD-ERGM}   &
\multicolumn{2}{c}{\rule{0pt}{15pt}  \% Correct Results   } \\
&$\theta_1\et$&$\theta_2\et$&$\theta_1\et$&$\theta_2\et$ & $\tonde{\theta_1\et \; , \theta_2 }$ & $\tonde{\theta_1 \; , \theta_2\et }$ \\
\hline
\rule{0pt}{12pt}  Const   & 0.02 & 0.1  &  0.0006 & 0.004 &&\\
\hline
\rule{0pt}{12pt}  Sin   & 0.02 & 0.04 &   0.003 & 0.005 &94\%&93\%\\
\hline
\rule{0pt}{12pt}  Steps  &  0.02 & 0.03 &  0.01 & 0.001 &92\%& 96\%\\
\hline
\rule{0pt}{12pt}  AR(1)  & 0.02 & 0.2 &  0.007 & 0.01 & 93\% &90\%\\
\end{tabular}
\end{table} 
It is worth noticing that, for sampling and cross-sectional inference, we employed the R package \textit{ergm} that uses state-of-the-art MCMC techniques for both tasks~\citep[see][for a description of the software]{hunter2008ergm}. Hence, we compared the SD-ERGM based on the approximate pseudo-likelihood -- both in the definition of the time-varying parameter update and inference of the static parameters -- with a sequence of exact cross-sectional estimates that are in general known to be better performing than the pseudo-likelihood alternative, as mentioned in Section \ref{sec:lit_rev_ergm}. Even if the cross-sectional estimates are based on the exact likelihood, while the SD approach is based on an approximation, the SD-ERGM remains the best-performing solution. This provides further evidence of the advantages of SD-ERGM as a filtering tool.
Finally, the last column of Table \ref{tab:mse_filter_gwesp} reports the percentage number of times the LM test of~\cite{calvori2017testing} applied to the SD-ERGM correctly classifies the parameters as time-varying (or static for the constant DGP). The test performs correctly in all the cases considered. 

\subsubsection{Comparison of Pseudo and Exact Likelihood SD-ERGM}\label{sec:pmle_mle_comparison}
To further investigate the proposed SD-ERGM and its version based on the pseudo-likelihood, in this section, we focus on the ERGM having the total number of links and the total number of mutual links as network statistics: 
\be\label{eq:p_star}
\sum_{s} \theta_s h_s\tonde{\Yt} = \theta_L \sum_{ij} \Yijt + \theta_M \sum_{ij} \Yijt  Y_{ji}\et.
\ee
The static version of this model is known as \textit{reciprocity} $p^\star$ model \citep{snijders2002markov}. This model is relevant for our discussion because it allows us to compare the SD time-varying extension based on the pseudo-likelihood with the one based on the exact likelihood. Indeed, it is simple enough that the normalizing function is known in closed form, but it has enough structure that its pseudo-likelihood differs from its exact likelihood. The model results in dyads, i.e., pairs of mutual links $(A_{ij}, A_{ji})$,  being independent, while the pseudo-likelihood amounts to assuming independent links. Moreover, since its partition function is available in closed form, such a model can be sampled efficiently without resorting to MCMC methods. This allows us to run extensive numerical simulations in reasonable time to investigate the properties of the confidence bands proposed by Ref.~\onlinecite{buccheri_etal2018smoother} in the context of SD-ERGM models.

In this section, we will refer to the pseudo-likelihood-based SD-ERGM as PML-SD-ERGM and to the exact likelihood case as ML-SD-ERGM. 
We compare the capacity of the two models, used as filters, to recover misspecified dynamics using the same approach as in the previous sections, i.e., we simulate a known DGP for $\theta_L\et$ and $\theta_M\et$. We focus on a DGP where $\theta_L$ and $\theta_M$ follow two independent $AR(1)$ processes, as the one discussed in  \ref{sec:beta_model}. Each $AR(1)$ has $\Phi_1  = 0.98$ and $\epsilon \sim N\tonde{0,\sigma}$ with $\sigma = 0.005$. The $\Phi_{0}$ parameters are chosen such that, on average, the network density equals $0.3$, and the fraction of reciprocated pairs is $0.075$. We select this value because it is between the maximum and minimum fraction of reciprocated links possible for a network of density $0.3$, $0$ and $0.3 (N^2-N)/2$, respectively. When comparing results for different network sizes, we keep the density fixed for all network sizes $N$, thus exploring a dense regime\footnote{We found the conclusions of this section to hold also in a sparse network density regime.}.
In our numerical experiment, we first sample sequences of synthetically generated observations repeatedly from different specifications of the DGP. We then estimate the PML and ML versions of the SD-ERGM on those observations and filter the tvps. Finally, we quantify their accuracy, with the average RMSE, across 50 samples with respect to the simulated DGP.  In Table \ref{tab:mse_filter_pstar}, we report the RMSE for both PML-SD-ERGM and ML-SD-ERGM, divided by the RMSE of the cross-sectional standard ERGM, for various combinations of network size $N$ and number of observations $T$. It emerges that both versions of SD-ERGM strongly outperform the cross-sectional ERGM. Moreover, the performances of PML-SD-ERGM are similar to the ones of the exact ML-SD-ERGM.

\begin{table} 
\caption{\label{tab:mse_filter_pstar} RMSEs of ML-SD-ERGM and PML-SD-ERGM, relative to that of the cross-sectional ERGM. The averages are obtained over 50 repetitions for the $AR(1)$ DGP described in the text. For each value of $T$ and $N$, we report the RMSE of the SD-ERGMs divided by the RMSE of the cross-sectional ERGM. } 
\centering
\begin{tabular}{|c||ccc|ccc|}
\hline
&   \multicolumn{3}{c|}{PML-SD-ERGM}   &  \multicolumn{3}{c|}{ ML-SD-ERGM}   \\
\hline
\rule{0pt}{12pt}  \diagbox{T}{N} & 50 & 100  &  500 & 50 & 100  & 500 \\
\hline
\hline
\rule{0pt}{12pt}  100   & 0.016 & 0.011  &  0.006 & 0.015 & 0.011& 0.006\\
\hline
\rule{0pt}{12pt}  300   & 0.015 & 0.011  &  0.006 & 0.014 & 0.011& 0.007\\
\hline
\rule{0pt}{12pt}  600   & 0.014 & 0.012  &  0.007 & 0.014 & 0.011& 0.006\\
\hline
\end{tabular}
\end{table} 

In the final part of this section, we investigate the possibility of using the method of Ref.~\onlinecite{buccheri_etal2018smoother}, that we describe in Appendix~\ref{sec:appendix_DCS}, to define confidence bands for the parameters filtered with SD-ERGM.
The authors characterize the approximation error when the SD approach filters a set of latent parameters whose true DGP is an auto-regressive process. While we refer to the original manuscript for the details, we point out that their procedure rests upon the assumption that the SD filter approximates the true underlying DGP. The authors prove that this approximation becomes exact in the limit of small variance for the latent parameters. Hence, the confidence bands obtained with their method are theoretically guaranteed to be reliable only in this limit. In practice, assessing whether the application of the confidence bands is justified for a given value of the variance of the DGP is appropriate. Numerical experiments can do this to determine their coverage with a simulated DGP. For example, for the model and the DGP considered in this section, we check the coverage of the confidence bands obtained and report the results in Table \ref{tab:mse_cov_pstar}, for $N=100$.
\begin{table} 
\caption{\label{tab:mse_cov_pstar} Coverages of the 95\%  confidence bands averaged over 50 repetitions, for the $AR(1)$ DGP, described in the text, and $N=100$. }  
\centering
\begin{tabular}{|c||c|c|}
\hline
$T$ & ML-SD-ERGM   &  PML-SD-ERGM   \\
\hline
\hline
\rule{0pt}{12pt}  300   & 99.1 \% & 99.9 \%  \\
\hline
\rule{0pt}{12pt}  3000   & 94.5\% & 95.7 \% \\
\hline
\end{tabular}
\end{table} 
We find that the coverage of the confidence bands, for both ML-SD-ERGM and PML-SD-ERGM, approaches the nominal value in the limit of large $T$, while for short time series, their coverages are higher than the nominal value. Hence, in small samples, they should be interpreted as having a confidence of \textit{at least} their nominal values.

\section{Applications to Real Data}\label{sec:applications}

After analyzing synthetic data, this section presents two applications to real temporal networks. Our goal is to show the value of SD-ERGM as a methodology to model temporal networks, irrespective of the specific system that a researcher wants to investigate. The two real networks that we consider have been the object of multiple studies in different streams of literature. They have been investigated in the context of ERGMs using different network statistics. We first consider a network of credit relations among Italian banks. The second real-world application focuses on a network of interest for the social and political science community, namely the network of U.S. senators cosponsoring legislative bills.

\subsection{Link Prediction in Interbank Networks}
Our first empirical application is to data from the electronic Market of Interbank Deposit (e-MID). In this market, banks can extend loans to one another for a specified term and/or collateral. Interbank markets are an important point of encounter for banks' supply and demand of extra liquidity. In particular, e-MID has been investigated in many papers
\cite{iori2008network,finger2013network,mazzarisi2020dynamic,barucca2018organization}.
Our dataset contains the list of all credit transactions on each day from June 6, 2009, to February 27, 2015. Our analysis investigates the interbank network of overnight loans aggregated weekly. We follow the literature and disregard the size of the exposures, i.e., the weights of the links. We thus consider a link from bank $j$ to bank $i$ present at week $t$ if bank $j$ lent money overnight to bank $i$, at least once during that week, irrespective of the amount lent. This results in a set of $T = 298 $ weekly aggregated networks. For a detailed dataset description, we refer the reader to Ref.~\onlinecite{barucca2018organization}.

In recent years, the amount of lending in e-MID has significantly declined. In particular,  it abruptly decreased at the beginning of 2012 due to important unconventional measures (Long Term Refinancing Operations) by the  European Central Bank that guaranteed an alternative source of liquidity to European banks. The evident non-stationary nature of the evolution of the interbank network is of extreme interest to us. As mentioned in Sections \ref{sec:beta_model} and \ref{sec:SDERGM_glob_Stat}, one of the key strengths of SD-ERGM, used as a filter, is precisely the ability to recover such non-stationary dynamics.

In the following, we use the SD beta model for link forecasting. Specifically, we consider the version with a restricted number of static parameters discussed at the end of Sec. \ref{sec:sim_sd_beta}. We divide the data set into two samples.
We consider rolling windows of $100$ observations and  estimate the parameters  $\alpha^{\text{out}}$, $\beta^{\text{out}}$, $\alpha^{\text{in}}$ and $\beta^{\text{in}}$ on each one of those rolling windows. For each window, we test the forecasting performances up to $8$  steps ahead (i.e., roughly two months). The forecast works as follows. Assuming that at time $t$, the last date of the rolling window, we have filtered the value for the parameters $ \ibinpar\et $ and $ \obinpar\et$, we plug the estimated static parameters and the matrix $\Yt$ in the SD update and compute the tvps $ \ibinpar\etp $ and $ \obinpar\etp$. From the latter, we readily obtain the forecast of the adjacency matrix 
$$
\mathbb{E}\quadre{\Y^{(t+1)}\vert \ibinpar\etp, \obinpar\etp}\,,
$$  
where $t+1$ is the first date of the test sample. The $K$-step-ahead forecast for the SD-ERGM model is obtained by simulating the SD dynamics up to $t+K$ $100$ times\footnote{It is worth stressing that the results become stable after $20$ simulations.}, thus obtaining  $\obinpar^{(t+K)}_n$ and $\ibinpar^{(t+K)}_n$ for $n=1,\dots,100$, and then taking the average of the expected adjacency matrices $\frac{1}{100}\sum_n \mathbb{E}\quadre{\Y^{(t+K)}\vert \ibinpar^{(t+K)}_n, \obinpar^{(t+K)}_n}$. Given the forecast values, we compute the rate of false positives and false negatives. Then, we drop the first element from the train set and add the first element of the test sample. We repeat the forecasting exercise, estimating the SD-ERGM parameters on the new train set and testing the performance of the new test sample. We name this procedure a rolling estimate and iterate it until the test sample contains the last eight elements of the time series. 

\begin{figure}
 	\includegraphics[width=0.45\linewidth 	]{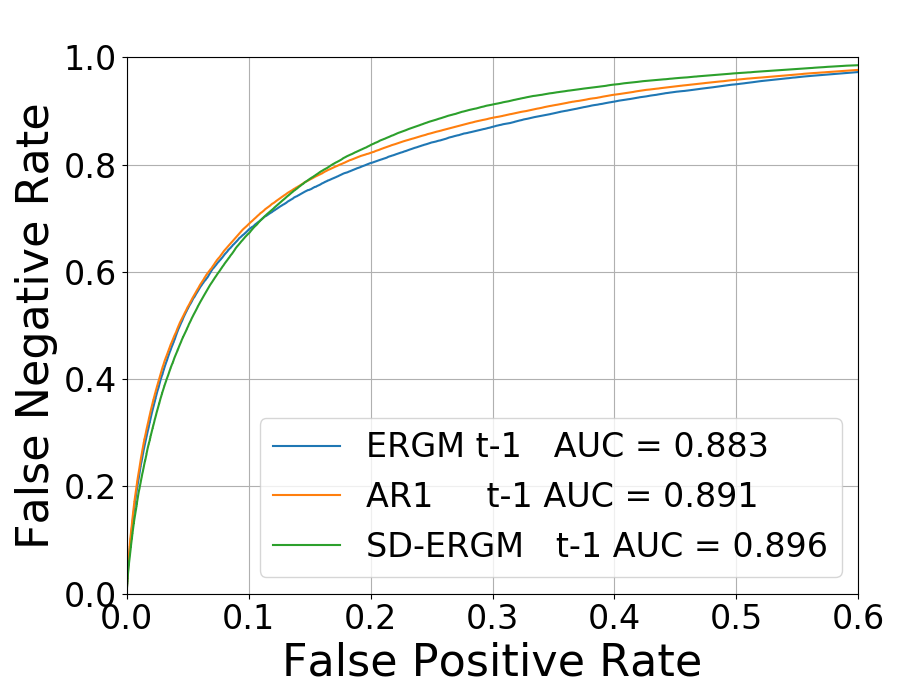}
 	\includegraphics[width=0.45\linewidth 	]{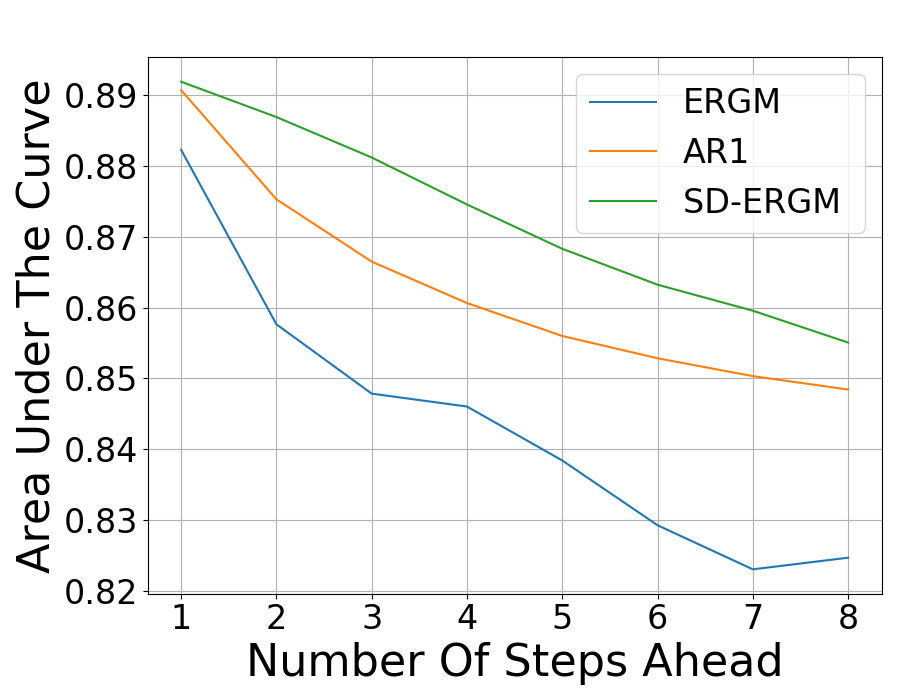} 
 	\caption{Left panel: ROC curves for one-step-ahead link forecasting.
 	The green and orange ROC curves describe the one-step-ahead forecasting with SD and cross-sectional AR(1) beta model, respectively. The blue curve corresponds to the forecast based on the ERGM at the previous time step. Right panel: AUC for the multi-step-ahead forecast.}\label{fig:emid_application_forecast}
\end{figure}

Given a forecast for the adjacency matrix, we evaluate the accuracy of the binary classifier by computing the Receiving Operating Characteristic (ROC) curve. All results are collected and presented in Fig.~\ref{fig:emid_application_forecast}. The left panel reports the ROC curve for one-step-ahead link forecasting obtained according to the SD-ERGM rolling estimate. The panel also shows three other curves based on the static beta model. Specifically, the green curve results from a naive prediction, where a link tomorrow is forecasted, assuming that the $t+1$ ERGM parameter values are equal to those estimated at time $t$. Once the sequence of cross-sectional estimates of the static ERGM is completed, we take the estimated values $ \widehat{\ibinpar}\et$  and $ \widehat{\obinpar}\et$ as observed and model their evolution with an auto-regressive model of order one, AR(1). That amounts to assuming $ \widehat{\ibinpar}\etp = c_0 + c_1 \widehat{\ibinpar}\et + \epsilon\et$, where $c_0$ and $c_1$ are the static parameters of the AR(1), and $\epsilon\et$ is a sequence of i.i.d. normal random variables with zero mean and variance $\sigma^2$. A similar equation holds for the out-degree parameters. Using the observations from the training sample, we estimate the parameters $c_0$, $c_1$, and $\sigma^2$ and use them for a standard AR(1) forecasting exercise on the test sample. The results correspond to the orange curve. It is important to stress that while the SD-ERGM forecast requires one static and one time-varying estimation on the train set, we must estimate the static parameters for each date in the train sample in the latter procedure. 

The left plot of Fig.~\ref{fig:emid_application_forecast} shows that the naive one-step-ahead forecast, despite its simplicity, provides a reasonable result. The best performance corresponds, however, to the forecast based on the SD-ERGM. The AR(1) static ERGM improves on the naive forecast and is slightly worse than the SD-ERGM. However, as commented before, it is more computationally intensive. More importantly, the right panel of Fig.~\ref{fig:emid_application_forecast} presents a multi-step-head forecasting analysis result. 
It emerges that the naive forecast's performance (blue curve), tested up to $K=8$, rapidly deteriorates. In contrast, the SD-ERGM multi-step forecast remains the best performing~\footnote{In all the results on link forecasting -- one- or multi-step-ahead -- we excluded the links that are always zero, i.e., they never appear in the train and test samples. The reason is that those are extremely easy to predict, and keeping them would give an unrealistically optimistic picture of the predictability of links in the data set. Notably, the ranking of the methods remains unaltered when we keep all links for performance evaluation.}.

\subsection{Temporal Heterogeneity in U.S. Congress Co-Voting Political Network}\label{sec:us_congr_app}

Networks describing the U.S. Congress' bills have been the object of multiple studies~\citep[see,  for example][]{fowler2006connecting,faust2002comparing,zhang2008community,cranmer2011inferential,moody2013portrait,wilson2019modeling,lee2020varying}. It is thus an appropriate real system for our second application of the SD-ERGM framework. In particular, we want to show that the update rule based on the pseudo-score defined in~\eqref{eq:pseudoscore} can be concretely applied to a real network and draws a different picture when compared to the sequence of cross-sectional ERGM estimates. To build the network, we use the freely available data of voting records in the US Senate~\citep[see ][]{voteview} covering the period from 1867 to 2015, for a total of 74 Congresses. We define the network of co-voting following Refs.~\onlinecite{roy2017change} and \onlinecite{lee2020varying}, where a link between two senators indicates that they voted in agreement on over 75\% of the votes among those held in a given senate when they were both present.  This procedure results in 74 networks, one for each Congress, starting from the $40$th.  
\begin{figure}
	\includegraphics[width=0.9\linewidth 	]{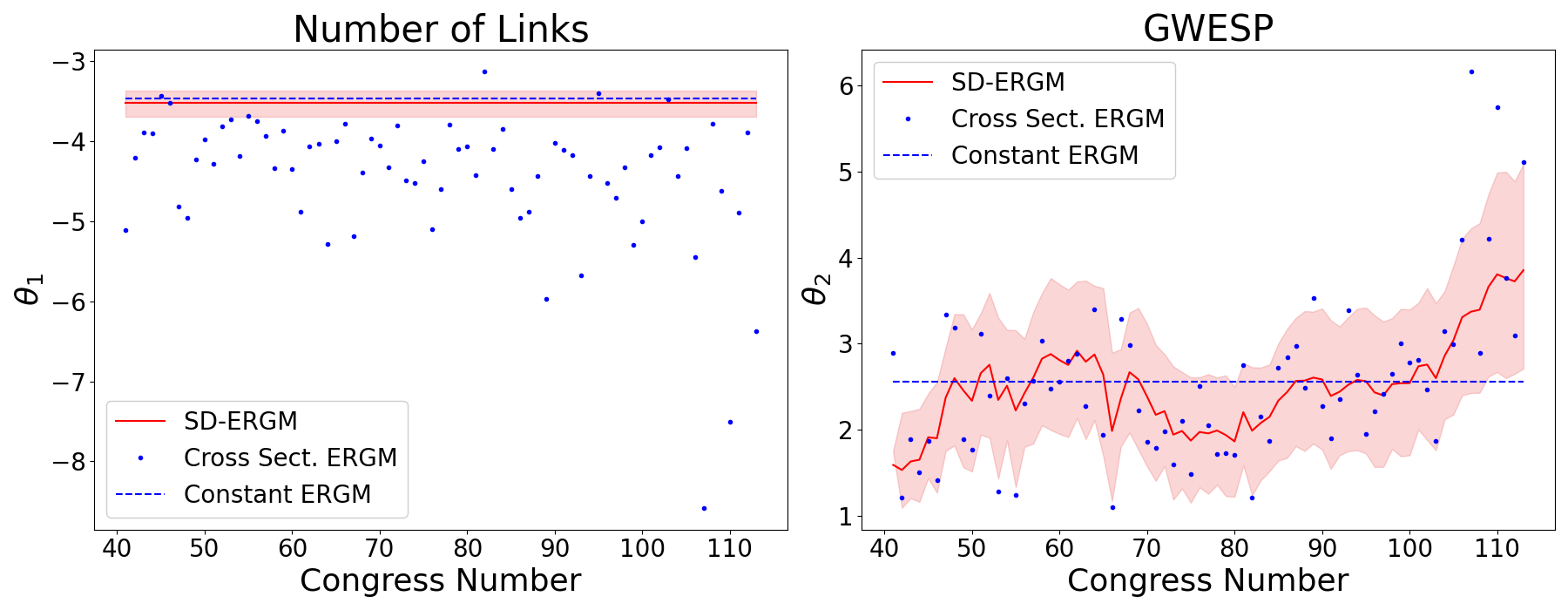} \\
	\caption{ Estimates for the tvps associated with the number of links and the GWESP statistics. Blue dots correspond to the cross-sectional ERGM estimates. The red lines are the estimates from the SD-ERGM, with the corresponding $95\%$ confidence intervals denoted by the red-shaded regions.}\label{fig:covoting_filter} 
\end{figure}
We consider the SD-ERGM with the two network statistics discussed in Section~\ref{sec:SDERGM_glob_Stat} for this empirical application. As defined in \eqref{eq:ERGM_glob_def}, parameter $\theta_1\et$ is associated with the number of edges, while $\theta_2\et$ with the GWESP statistic. 
The fact that the number of nodes is not constant over time is not a problem for our application since we do not consider statistics associated with single nodes. That case -- as, for instance, considering the degrees of the beta model -- would require the number of tvps to be different at each time step.  

As we did for the numerical simulations and the previous empirical application, we compare our framework with a sequence of standard ERGMs. This empirical exercise does not aim to conclude the specific network at hand. We aim to show that the two approaches return a qualitatively different picture. The choice between the alternative models and combinations of statistics—possibly based on model selection techniques—is beyond the scope of our exercise.

Using the test for temporal heterogeneity based on SD-ERGM, only the parameter $\theta_2$ turns out to be time-varying. Testing the null hypothesis that each parameter is static, we obtain a p-value of $0.1$ for the link density and  $10^{-4}$ for GWESP. To check whether the sequence of cross-sectional estimates is consistent with the hypothesis that the parameters remain constant, we estimate the values $\theta_1^c$, $\theta_2^c$ from an ERGM using all observations. This amounts to compute $\theta^c = \argmax{\theta} \sum_{t=1}^{74} \log P\tonde{\Y\et,\theta}$. Then, for each sequence of cross-sectional estimates $\theta_1\et$ and $\theta_2\et$, we test the hypothesis of them being normally distributed around the constant values with unknown variance. The p-values resulting from the \textit{t}-tests are  $1.4\times 10^{-6}$ and $0.03$ for parameters $\theta_1$ and $\theta_2$, respectively.  This simple test confirms that the two approaches imply quantitatively different parameter behaviors. This emerges from Fig.~\ref{fig:covoting_filter} that reports the estimates from the SD-ERGM (thick red lines), with their respective 95\% confidence intervals (shaded red bands), as well as the cross-sectional ERGM estimates -- one per date (blue dots) or using the entire sample (dashed blue line).

To compute the confidence bands as in Ref.~\onlinecite{buccheri_etal2018smoother}, we numerically checked whether the data is compatible with a DGP with a small variance. In practice, we first estimate the SD-ERGM. Then we quantify the variance of the latent parameters by estimating an AR(1) on the filtered time series\footnote{We extensively tested via simulation that, for the model at hand and $T$ and $N$ taken from the data, estimating the variance of the latent parameters in such a way results, on average, in a small underestimation. In checking the coverage of the confidence bands, we considered a DGP with variance increased, with respect to the one estimated on the filtered time series, to compensate for this bias.}. Finally, we repeatedly simulate such an $AR(1)$ DGP, similarly to what is done at the end of section \ref{sec:pmle_mle_comparison}, and check the coverage of the confidence bands. We find that, for the current application, the coverage of the confidence bands is 99.9\%, hence larger than the nominal value. These simulation-based results support the reliability of the approximate SD filter and provide a conservative estimate of the confidence bands. This allows us, for example, to deduce that the data is not compatible with a model where one of the two parameters is zero.

\section{Conclusions}\label{sec:conclusions}
In this paper, we proposed a framework for describing temporal networks that extends the well-known Exponential Random Graph Models. In the new approach, the parameters of the ERGM have stochastic dynamics driven by the conditional likelihood score. If the latter is unavailable in closed form, we showed how to adapt the score-driven updating rule to a generic ERGM by resorting to the conditional pseudo-likelihood. In this way, our approach can describe the dynamic dependence of the PMF from virtually all the network statistics usually considered in ERGM applications. We investigated two specific ERGM instances using an extensive Monte Carlo analysis of the SD-ERGM reliability as a filter for tvps. The chosen examples allowed us to highlight the applicability of our method to models with a large number of parameters and to models for which the normalization of the PMF is not available in closed form. The numerical simulations proved the clear superior performance of the SD-ERGM over a sequence of standard cross-sectional ERGM estimates. This is true not only in the sparse network regime but also in the dense case when the number of nodes is far from the asymptotic limit. Finally, we run two empirical exercises on real network data. The first application to the e-MID interbank network showed that the SD-ERGM provides a quantifiable advantage in a link forecasting exercise over different time horizons. The second example of the U.S. Congress co-voting political network enlightened that the ERGM and the SD-ERGM could provide a significantly different picture describing the parameter dynamics. 

Our work opens several possibilities for future research. First, the applicability of the test for parameter instability in the context of SD-ERGM with multiple network statistics could be investigated much further. This would require an in-depth analysis of the multi-collinearity issues intrinsic to the ERGM context. 
Second, the SD-ERGM could be applied to multiple instances of real-world temporal networks. An interesting application would be the study of networks describing the dynamical correlation of neural activity in different parts of the brain~\cite{karahanouglu2017dynamics}. In this context, applying the static ERGM has already proven highly successful~\cite{simpson2011exponential}.
The last future development we plan to explore is extending the score-driven framework to the description of weighted temporal networks. Regretfully, this setting deserves more attention in the literature~\cite{giraitis2016estimating}. Still, it is of extreme relevance, particularly from the financial stability perspective and its implications for systemic risk.

\begin{acknowledgments}
We are particularly thankful for the comments and suggestions received by Fulvio Corsi and Giuseppe Buccheri. Fabrizio Lillo acknowledges partial support by the European Program scheme ’INFRAIA-01-2018- 2019: Research and Innovation action’, grant agreement \#871042 ’SoBigData++: European Integrated Infrastructure for Social Mining and Big Data Analytics.’ Giacomo Bormetti and Fabrizio Lillo acknowledge financial support from the Italian Ministry MUR under the PRIN project \textit{Dynamic models for a fast changing world: An observation driven approach to time-varying parameters} (grant agreement no. 20205J2WZ4).
\end{acknowledgments}

\section*{Author Declarations}

\subsection*{Conflict of Interest}

The authors have no conflicts to disclose.

\section*{Data Availability Statement}

We are not allowed to share the data from the e-MID inter-bank market, as a confidentiality agreement between the Scuola Normale Superiore of Pisa and the data provider, LIST SpA, binds us. The interested reader can contact \href{https://www.list-group.com/}{LIST SpA} for inquiries concerning data access.
The second application's data are publicly available at \url{https://voteview.com/data}.

\appendix

\section{Exponential Random Graphs Models}\label{sec:appendix_ERGMs}
Here, we review some standard notions regarding ERGMs for the reader's convenience.  A PMF over the set of graphs as in eq.~\eqref{eq:ergm_def} defines an ERGM.
To begin with, let us mention the first and probably most famous example of this class: the Erd\"os-R\'enyi model~\cite{erdds1959random}.  In this model, for a given number of nodes $N$, each of the possible $N\tonde{N -1 }/2$ links~\footnote{In the whole paper, we do not allow for links that start and end at the same node, so named \textit{self-loops}. However, including them would be trivial.} is present with constant probability $p$, equal for all links. The probability to observe the adjacency matrix $\Y$ is 
\begin{equation*}
	P\tonde{\Y} = \prod_{i< j} p^\Yij \tonde{1-p}^{\tonde{1-\Yij}}.
\end{equation*}  
In the context of exponential distributions,  it is possible to consider more general structures for the probability of a link to be present and even depart from the assumption that each link is independent of the others.  Examples of more general ERGMs have been first proposed in Ref.~\onlinecite{Holland81anexponential}, under the name of log-linear, or $p^\star$, models. For instance, the $p1$ model is defined by the PMF
\begin{equation*}
\label{p1_def}
\log{P\tonde{\Y}} = \sum_{ij}\quadre{ Y_{ij} Y_{ji} \rho_{ij} +  Y_{ji} \phi_{ij} } - \log{\tonde{\mathcal{K}\tonde{\boldsymbol{\rho},\boldsymbol{\phi}}}}, 
\end{equation*} 
where $\boldsymbol{\rho}$ and $\boldsymbol{\phi}$ are two matrices of parameters, and $\mathcal{K}\tonde{\boldsymbol{\rho},\boldsymbol{\phi}}$ is a normalization factor, also known as partition function in the statistical physics literature.  
This model can be estimated in parsimonious specifications, e.g., $\phi_{ij} = \phi_i +\phi_j$, known as \textit{sender plus receiver effect}, and $\rho_{ij} = \rho$ that describes the tendency to reciprocate links. Additionally, p1 models can be enriched with dependencies on node attributes \citep[][]{fienberg1981categorical} or predetermined (exogenous or endogenous) covariates $X_{ij}$~\citep{wasserman1996logit}. The requirement of independence among dyads has been relaxed since Ref.~\onlinecite{frank1986markov} to take into account neighborhood effects, such as the tendency to form 2 stars, quantified by the function $h_{2\text{-stars}} = \sum_{ijk} Y_{ik} Y_{jk}$ or triangles $h_{\text{triangles}} = \sum_{ijk} Y_{ik} Y_{kj} Y_{ji} $. These functions are examples of network statistics, i.e., functions of the adjacency matrix, that play a central role in ERGMs.

\subsection{The Beta Model }\label{sec:beta_model_intro}

The first example we consider is quite simple but, at the same time, largely employed in different streams of literature \citep[][]{zermelo1929berechnung,Holland81anexponential,PhysRevLett.89.258702,PhysRevE.70.066117,PhysRevE.78.015101,chatterjee2011random}. The range of applications for this model is so broad that researchers were often not aware of previous works using the same model. For this reason, it can be found under at least three different names: \textit{beta model}, \textit{fitness model}, and \textit{configuration model}. They all refer to a probability distribution that can be rewritten as an ERGM where each node $i$ has two parameters: $\obinpar_i$, that captures the propensity of node $i$ to form outgoing connections and $\ibinpar_i$ those incoming. It is standard to indicate the number of connections a node has as its \textit{degree}. For the directed network case considered here, we have -- for node $i$ -- \textit{out-degree} $\overrightarrow{D}_i$ and \textit{in-degree} $\overleftarrow{D}_i$ defined as
\be
\overrightarrow{D}_i = \sum_j \Yij\;,  \qquad \overleftarrow{D}_i = \sum_j Y_{ji}\,. \nonumber  
\ee
With these definitions, and since it is possible to compute the normalization factor $\mathcal{K}\tonde{\ibinpar, \obinpar}$, the PMF reads
\be \label{eq:beta_model}
\log{P\tonde{\Y}} =   \sum_{i=1}^N \left(\ibinpar_i \overleftarrow{D}_i   + \obinpar_i \overrightarrow{D}_i \right)-  \sum_{ij} \log{\tonde{1 + e^{\ibinpar_i + \obinpar_j}} }.
\ee
In the main text, we extensively use the beta model's score to define its score-driven extension. Hence, we show its explicit expression here for the reader's convenience. Defining, for ease of notation, 
$$
p_{ij} = \frac{1}{1+ e^{ - \ibinpar_i - \obinpar_j}} 
$$
we can write the score as  
$$
\nabla\tonde{ \ibinpar , \obinpar }   = \left(\begin{array}{c}
\frac{\partial  \log{P\tonde{\Y |  \ibinpar , \obinpar }} }{\partial \ibinpar}\\
\\
\frac{\partial  \log{P\tonde{\Y |  \ibinpar , \obinpar }} }{\partial \obinpar}
\end{array} \right) = \left(\begin{array}{c}
\sum_{i} \left(Y_{i1} - p_{i1}\right)\\
\vdots\\
\sum_{i} \left(Y_{iN} - p_{iN}\right)\\
\\
\sum_{i} \left(Y_{1i} - p_{1i}\right)\\
\vdots\\
\sum_{i} \left(Y_{Ni} - p_{Ni}\right)\\
\end{array} \right)
$$

The beta model is often used when the degree heterogeneity is expected to play a prominent role in explaining the presence or absence of links. It is worth noticing that the static version of the beta model, in the directed case, is not identified. Indeed, the probabilities remain unchanged after the application of the following transformation
$$
\left\lbrace\begin{array}{lcr} \ibinpar &\rightarrow &\ibinpar + c  \\
&&\\
\obinpar &\rightarrow &\obinpar - c\,.
\end{array} 
\right.
$$
The issue can be tackled \citep{yan2016asymptotics} by choosing one identification restriction that eliminates the possibility of shifting all parameters by an arbitrary constant. This is essential to compare the parameter values estimated for the same network at different times. In all our investigations, both the numerical simulations and the empirical applications, we enforce the following condition: 
$$\sum_i \ibinpar_i = \sum_i \obinpar_i. $$
It is worth noticing that different choices are available, e.g., $\sum_i \ibinpar = 0$ or $ \obinpar_i = 0$. However, and most importantly, the results presented in the paper do not change significantly when the identification condition changes. It is also important to notice that the MLE can be performed using a fixed point algorithm, described for example in \cite{yan2016asymptotics}, that reaches the optimal solution quickly. Moreover, we point out interesting results on the asymptotic behavior of the maximum likelihood estimates for $\tonde{\ibinpar, \obinpar } $ when the number of nodes increases. Indeed, consistency results have been proved in Ref.~\onlinecite{chatterjee2011random} for the undirected case and in Ref.~\onlinecite{yan2016asymptotics} for the directed case~\cite{yan2018statistical,jochmans2018semiparametric}. A necessary condition for these results to hold is that the network density remains constant as $N$ increases. An alternative, and often more realistic, possibility is that the average degree remains constant when $N$ increases, implying that the density decreases as\footnote{For a network with $N$ nodes, the number of possible links is of order $N^2$. Instead, when all nodes have a fixed average degree $d$, the number of present links is $d N$, and the density is of order $1/N$.} $1/N$. Networks belonging to this density regime are named \textit{sparse}. Notably, to our knowledge, no consistency results are known for large $N$ in the sparse regime.

\subsection{GWESP Statistic and Pseudo-Likelihood Estimation}
Here, we give some details on an example of a network statistic for which we cannot compute the partition function and the associated inference procedures. We focus on the GWESP. 

It is well known that when network statistics involve products of matrix elements\footnote{Examples of such statistics are the count of 2 stars present in the network or the number of triangles ~\citep{wasserman1996logit}.}, this is often the case. This lack of analytical tractability has arguably been the main obstacle in estimating ERGMs and understanding their properties. Moreover, it is nowadays well known that, when dealing with ERGMs, the use of network statistics involving products of matrix elements, such as the number of triangles, requires some care to avoid statistical issues  \citep{handcock2003assessing,handcock2003statistical}. The main problem, with consequences on estimation, simulation, and interpretability of ERGMs,  is {\it degeneracy}. An ERGM is degenerate if it concentrates a significant portion of its probability on a small set of configurations, typically the uninteresting graphs completely connected or void of links. When this phenomenon occurs, estimating the model becomes very hard, and often, the estimated model does not provide a meaningful description of real networks.  A great effort has been dedicated to investigating this problem and characterizing degeneracy~\cite{schweinberger2011instability}. A family of network statistics that, while describing properties of the whole network, is not plagued by degeneracy has been proposed in Refs.~\onlinecite{snijders2006new,robins2007recent} and discussed in Ref.~\onlinecite{hunter2006inference}. This family is called curved exponential random graphs, and one example of curved statistics is the GWESP. This function has recently been applied extensively to describe transitivity in social networks~\citep[see][]{hunter2006inference}. It captures the tendency of nodes to form triangles without the degeneracy issues that emerge when the direct triangle count is used as a statistic in ERGM. To get an intuition of the formula defining GWESP, let us consider two nodes connected by an edge and count the number of nodes to which they are connected, i.e., the number of neighbors they share. Let us indicate with $\text{ESP}_{k}\tonde{\Y}$  the number of edgewise shared partners, i.e., connected node pairs~\footnote{\textit{Edgewise} precisely means that we count partners only if shared by nodes that are connected.} that share exactly $k$ neighbors in the network described by $\Y$. Then GWESP is defined as 
$$ 
\text{GWESP}\tonde{\Y} =  e^{\lambda} \sum_{k=1}^{n-2} \quadre{1 - \tonde{1- e^{-\lambda}}^k } \text{ESP}_{k}\tonde{\Y}\,.
$$
In the following, we will be stuck to the usual approach in the literature, treating the parameter $\lambda$ as fixed and known, i.e., $\lambda = 0.5$. 

As mentioned in the main text, there are two standard approaches
to ERGM inference when the partition function is not available in closed form.
The first possibility~\citep{snijders2002markov} is to maximize an objective function obtained from a sufficiently large sample drawn from the PMF with an arbitrary (but close enough to the true one) parameter. As a consequence of the non-independence of the links in the general ERGM, sampling from \eqref{eq:ergm_def} necessary relies on Markov Chain Monte Carlo (MCMC) approaches \citep[see ][for a description of a popular software that implements it]{hunter2008ergm}. The computational burden of MCMC-based estimation can be prohibitive for graphs that are large enough. For this reason, a second approximate inference procedure, known as Maximum Pseudo-Likelihood Estimation (MPLE), first proposed for ERGMs in the seminal work of Ref.~\onlinecite{strauss1990pseudolikelihood}, is often used in empirical applications. 
MPLE is based on optimizing the pseudo-likelihood function, defined from link-specific variables (one for each element of the adjacency matrix) named \textit{change statistics}. Given an ERGM, the change statistic for the link between node $j$ and $i$, associated with network statistic $h_s$ is $\delta^s_{ij} = h_s\tonde{\Y_{ij}^+} - h_s\tonde{\Y_{ij}^-}$, where $\Y_{ij}^+$ is a matrix such that $Y^+_{ij} = 1$ and it is equal to $\Y$ in all other elements. Similarly, $\Y_{ij}^-$ has $Y^-_{ij} = 0$ and it is equal to $\Y$ in all other entries. Given these definitions, the pseudo-likelihood reads
\begin{equation}\label{eq:pseudolikelihood_def}
PL\tonde{\Y} = \prod_{ij} \pi_{ij}^{\Yij} \tonde{1-\pi_{ij}}^{\tonde{1-\Yij}} 
\end{equation} 
where $\pi_{ij} =\tonde{1+ e^{-\sum_s \theta_s \delta^s_{ij}}}^{-1}$. 

Pseudo-likelihood inference is of crucial importance when applying our methodology to any ERGM. Obtaining the pseudo-likelihood estimates is much faster than the MLE based on MCMC and easy to implement since the pseudo-likelihood boils down to the likelihood of a logistic regression. Then, it can be efficiently maximized with standard software for logistic regressions. However, the analogy with logistic regression is typically pushed too far. It has become widespread malpractice to associate with MPLEs the confidence intervals obtained from the maximum-likelihood theory for logistic regressions that are known to be theoretically unjustified, as already noted in Refs.~\onlinecite{strauss1990pseudolikelihood} and \onlinecite{handcock2003statistical}, and thoroughly discussed in Ref.~\onlinecite{varin2011overview}.
It is nowadays common knowledge that such a naive approach to MPLE inference results in a systematic underestimation of confidence intervals' width~\cite{vanduijn2009, desmarais2012statistical,schmid2017exponential}. More principled methods to estimate uncertainties of MPLEs, based on non-parametric and parametric bootstrap, have been proposed in Refs.~\onlinecite{desmarais2012statistical} and \onlinecite{schmid2017exponential}, respectively. These contributions showed that the computational convenience of MPLE for ERGMs can be reconciled with a reliable estimation of statistical uncertainties.

\section{Score-Driven Models}\label{sec:appendix_DCS}
Several are the reasons for the flexibility of a score-driven approach and its success in time-series modeling. Here, we review some key concepts mentioned in the main text that might prove useful to readers unfamiliar with the relative literature.

Score-Driven models have been introduced as Dynamic Conditional Score models by Ref.~\onlinecite{harvey2013dynamic} and Generalized Autoregressive Score models by Ref.~\onlinecite{creal2013generalized}.
Given a sequence of observations $\graffe{y\et}_{t=1}^T$, where each $y\et \in\mathbb{R}^M$,  and a conditional probability density $P\tonde{y\et\vert f\et}$, depending on a vector of tvps $f\et \in \mathbb{R}^K$, a Score-Driven model assumes that the time evolution of $f\et$ is ruled by the recursive relation~\eqref{eq:gasupdaterule}.

In practical applications, the static parameters of \eqref{eq:gasupdaterule} must be estimated. As detailed in Ref.~\onlinecite{harvey2013dynamic}, the likelihood of Score-Driven models can be readily expressed in closed form using the so-called prediction error decomposition. In a univariate setting, Ref.~\onlinecite{blasques2014maximum} works out the required regularity conditions, ensuring the consistency and asymptotic normality for the maximum likelihood estimators of the parameter values.

There are motivations, originating in information theory, for the optimality of the score-driven updating rule. In Ref.~\onlinecite{blasques2015information}, the authors consider a true and unobserved DGP $y\et \sim P\tonde{y\et\vert f\et}$. They assume a given and, in general, misspecified conditional observation density $ \tilde{P}\et = \tilde{P}\tonde{ \;.\;\; \vert \tilde{f}\et}$,  and consider the Kullback-Leibler (K-L) divergence
$$
\mathcal{D_{KL}}\tonde{P\et,\tilde{P}\etp} =  \int_A   P\tonde{y\vert f\et} \log\frac{ P\tonde{y\vert f\et}}{\tilde{P}\tonde{ y \vert \tilde{f}\etp} } \;dy,
$$
where $A\subseteq\mathbb{R}$. Building on the minimum discrimination information principle \citep{kullback1997information}, they argue that when the new observation $y_t$ becomes available, $\tilde{f}\etp$ should ideally be such that the updated density $\tilde{P}\etp$ is as close as possible to the true density $P\et$. Given that the real DGP is unknown, an optimal update that minimizes $\mathcal{D_{KL}}$  cannot be defined in practice. For this reason, \cite{blasques2015information} focus on the improvements of $\mathcal{D_{KL}}$ that an updating step produces irrespectively of the true DGP. One way of quantifying the improvement for a parameter update from $\tilde{f}\et$ to $\tilde{f}\etp$ is to consider the realized variation of $\mathcal{D_{KL}}$ 
\begin{eqnarray*}
\Delta_{t\vert t} & \equiv \mathcal{D_{KL}}\tonde{P\et,\tilde{P}\etp} - \mathcal{D_{KL}}\tonde{P\et,\tilde{P}\et} \nonumber\\
 & =  \int_A P\tonde{y\vert f\et} \log\frac{ \tilde{P}\tonde{y\vert \tilde{f}\et}}{\tilde{P}\tonde{ y \vert \tilde{f}\etp} } \;dy\,.
\end{eqnarray*}
Based on this definition, a parameter update is realized K-L optimal when $\Delta_{t\vert t}<0$ for every $\tonde{y\et,\tilde{f}\et,f\et}$. 
The authors prove that, under reasonable assumptions, the updating rule \eqref{eq:gasupdaterule} based on the score of $\tilde{P}\etp$ is locally realized K-L optimal. For more details and alternative definitions of optimality, we direct the reader to the original work and the more recent Ref.~\onlinecite{blasques2020finite}. For our definition of the SD-ERGM, we want to stress that realized optimality defines
a class of updates; it does not represent a single update with a unique functional form. For instance,  $\Delta_{t\vert t}$ defined above is specific to the chosen $\tilde{P}$. A different choice of $\tilde{P}$, e.g., one inspired by the pseudo-likelihood specification, translates into an alternative optimal choice for the update. In general, there can be infinite realized Kullback-Leibler optimal updates.
We remark that from the information-theoretic perspective of Ref.~\onlinecite{blasques2015information}, an update based on the pseudo-score, as we propose in the main text, is not only admissible but also realized K-L optimal, i.e., at each step, it diminishes the K-L distance of the pseudo-PMF, which assume independence of links, from the PMF of the true and unobserved DGP.

Any filtering tool should provide an estimation of the uncertainty and confidence bands for the estimates.
Ref.~\onlinecite{blasques2016confbands} discussed methods to quantify the uncertainty associated with the Score-Driven filters when the DGP is a Score-Driven model. Specifically, they proposed a simulation-based method to define in-sample confidence bands around the filtered tvps. Their procedure starts from the maximum likelihood estimate of the static parameters, given observations $\graffe{y^{\tonde{t^\prime}}}_{t'=1}^{t-1}$. Given the MLE estimate, the method prescribes to repeatedly sample new parameters $\tonde{w,\boldsymbol{\beta},\boldsymbol{\alpha}}_i$ from a multivariate normal, centered around the MLE estimates, and variance-covariance matrix estimated with the Huber-White estimator \citep{huber1967behavior, white1980heteroskedasticity}. Then one uses each sample to filter a different sequence of tvps, from the same time series of observations, thus obtaining a sample of filtered paths $ \hat{f}_i\et =  \hat{f}_i\et  \tonde{w_i,\boldsymbol{\beta}_i,\boldsymbol{\alpha}_i,\graffe{y^{\tonde{t^\prime}}}_{t^\prime=1}^{t-1}}$ for $i=1,\dots, K$, where $K$ is the number of samples. Finally, each time $t$, one uses the obtained distribution $\hat{f}_i\et$ to calculate the appropriate percentiles defining the confidence bands. 
While this construction is intuitive and easy to implement in practice, it is meant to capture only the uncertainty due to the estimation of the static parameters, often referred to as \textit{parameter uncertainty}. Hence, the confidence bands reliably quantify uncertainty only when the DGP is score-driven. In other words, these bands do not consider what is known as \textit{filtering uncertainty}. This is the uncertainty because, in general, we do not know the true DGP and the score-driven filter may be regarded only as an approximate filter. 
Recently, Ref.~\onlinecite{buccheri_etal2018smoother} investigated the approximation error made by applying a score-driven filter to a time-varying parameter model following a different DGP. They found that, for a class of DGPs where the parameters follow an auto-regressive process, the approximation becomes exact in the limit of a small variance of the latent parameters. Moreover, they proposed a method to define confidence bands, inspired by Ref.~\onlinecite{HAMILTON1986387}, that accounts for filtering and parameter uncertainty in Score-Driven filters. While we refer to their paper for the derivation details, here we briefly describe the key steps of the procedure. The total conditional variance of the latent parameters is decomposed as the sum of two terms. One term captures the parameter uncertainty similarly to the approach of Ref.~\onlinecite{blasques2016confbands}. The other term captures the filtering uncertainty and can be written in terms of the static parameters $\tonde{w, \boldsymbol{\beta}, \boldsymbol{\alpha}}$ and the scaling matrix $S\et$ from \eqref{eq:gasupdaterule} as $P\et = \boldsymbol{\beta}^{-1} \boldsymbol{\alpha} \boldsymbol{S}\et$. In practice, the procedure consists of sampling $\tonde{w,\boldsymbol{\beta},\boldsymbol{\alpha}}_i$ and obtaining a distribution of filtered paths, as in \cite{blasques2016confbands}. Then for each time step $t$ the variance of the latent parameters is obtained as $\frac{1}{K}\sum_i \tonde{\hat{f}\et_i - \hat{f}\et}^2 + \frac{1}{K}\sum_i \boldsymbol{\beta}_i^{-1}\boldsymbol{\alpha}_i S_i\et $, where $\hat{f}\et$ is the path filtered using the maximum likelihood estimates. 

Finally, we review the main idea behind the test for temporal parameter variation of Ref.~\onlinecite{calvori2017testing}. The method consists of a Lagrange Multiplier (LM) test for the parameter $\alpha$ that multiplies the score in the one-dimensional version of the recursion \eqref{eq:gasupdaterule}. The null hypothesis $H_0$ is that the parameter $f\et$ is static, i.e., $\beta = \alpha=0$, corresponding to $w$. As explained in Ref.~\onlinecite{davidson2004econometric}, the LM statistic for the hypothesis $H_0$, versus the alternative $\alpha = \beta\neq 0$, can be conveniently obtained from an auxiliary regression. To allow for a coefficient $\beta$ different from $\alpha$, one can use the same arguments as in Ref.~\onlinecite{lee1991lagrange}. As discussed in Ref.~\onlinecite{calvori2017testing}, the LM statistic can be written as the explained sum of squares from the regression 
$$
\mathbf{1} = c_w \nabla_w\et + c_\alpha S\etm \nabla_w\etm \nabla_w\etprime + \text{residual},
$$
where $c_w$ and $c_\alpha$ are regression coefficients that can be estimated with any statistical software. It is worth noticing that, under the null, the score of the conditional density with respect to $f\et$ is equal to the score with respect to $w$. From standard asymptotic theory, it follows that the LM statistic is distributed as a $\chi^2$ with one degree of freedom. For a detailed test description, we refer the reader to Ref.~\onlinecite{calvori2017testing}. 

\section{Details of Numerical Simulations}\label{sec:appendix_num_details}
The main text refers to a set of DGPs used for numerical simulations. Although the different numerical experiments that we presented differ in the meaning and number of parameters, in every experiment, each of the parameters can be constant or evolve according to one of the following dynamical DGPs:
\begin{itemize}
	\item abrupt change of half the parameters at $t = T/2$, i.e., for odd $s$ we have $\overline{\theta}_s\et = \overline{\theta}_{1_s} $ for $t\leq T/2$ and  $\overline{\theta}_s\et = \overline{\theta}_{2_s} $ for $t > T/2$,  while for even $s$ it is  $\overline{\theta}_s\et = \overline{\theta}_{0_s} $ for $t=1\dots T$;
	\item smooth periodic variation for half the parameters, i.e., for odd $s$ we have $\overline{\theta}_s\et = \overline{\theta}_{0_s} + \tonde{\overline{\theta}_{2_s} - \overline{\theta}_{1_s}  } \sin\tonde{4\pi t/T + \phi_s} $ for $t=1\dots T$, where the $\phi_s$ are randomly chosen for each node, while  for even $s$ it is  $\overline{\theta}_s\et = \overline{\theta}_{0_s} $ for $t=1\dots T$;
	\item autoregressive of order $1$ (AR(1)), i.e., for odd $s$ we have $\overline{\theta}_s\et =  \Phi_{0_s} + \Phi_1 \overline{\theta}_s\etm + \epsilon\et  $ for $t=1\dots T$, where $\Phi_1  = 0.99$, $\Phi_{0_s}$ is chosen such that the unconditional mean is equal to $\theta_{0_s}$, $\epsilon \sim N\tonde{0,\sigma}$ and $\sigma = 0.1$. As in the previous cases, for even $s$ we keep $\overline{\theta}_s\et = \overline{\theta}_{0_s} $ for $t=1\dots T$.
\end{itemize}

The dynamics considered are such that element $s$ of vector $\theta$ remains bounded between $\theta_{1_s}$ and $\theta_{2_s}$. The values of $\theta_{1}$ and $\theta_{2}$ are fixed to allow fluctuations in the in and out degrees of the nodes, as follows.
The vector $\overline{\theta}_{0}$ is obtained by first generating two-degree sequences (in and out) such that the degrees linearly interpolate between a minimum degree $D_m = 3$ and a maximum of $D_M=8$. Then, we need to ensure that the degree sequence is graphicable, i.e., such that it exists one matrix of zeros and ones from which it can be obtained. We iteratively match links that make up the out-degree sequence with those that make up the in-degree sequence, starting with the largest in- and out-degrees. In practice, we start with an empty matrix, select the largest out-degree, and set the matrix element between this node and the node with the largest in-degree to one. If, at some point, we cannot entirely allocate a given out-degree, we disregard the leftover links outgoing from that node and move to the next one. This procedure amounts to populating the adjacency matrix until no more links can be allocated. The degree sequence associated with this adjacency matrix is guaranteed to be graphicable. The numerical values of $\overline{\theta}_{0}$ follow from the estimation of the static beta model. Finally, to gain additional heterogeneity in the amplitude of the fluctuations, we define $N$ values evenly spaced between $0.4$ and $1$, i.e., $c_s$ for $s=1\dots N$. We use them to define 
\ba
\overline{\theta}_{1_s} = \overline{\theta}_{0_s} + c_s \tonde{\overline{\theta}_{0_{s+1}} - \overline{\theta}_{0_s}  } \nonumber \\
\nonumber \\
\overline{\theta}_{2_s} = \overline{\theta}_{0_s} - c_s \tonde{\overline{\theta}_{0_{s+1}} - \overline{\theta}_{0_s}  } . \nonumber
\ea 

\begin{figure}
	\includegraphics[width=0.45\linewidth 	]{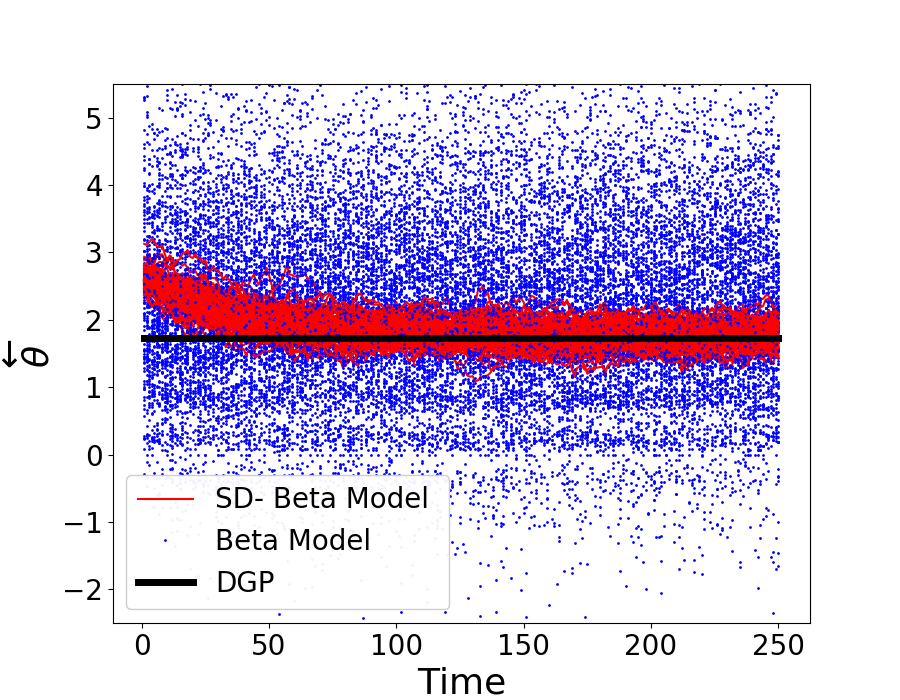}
	\includegraphics[width=0.45\linewidth 	]{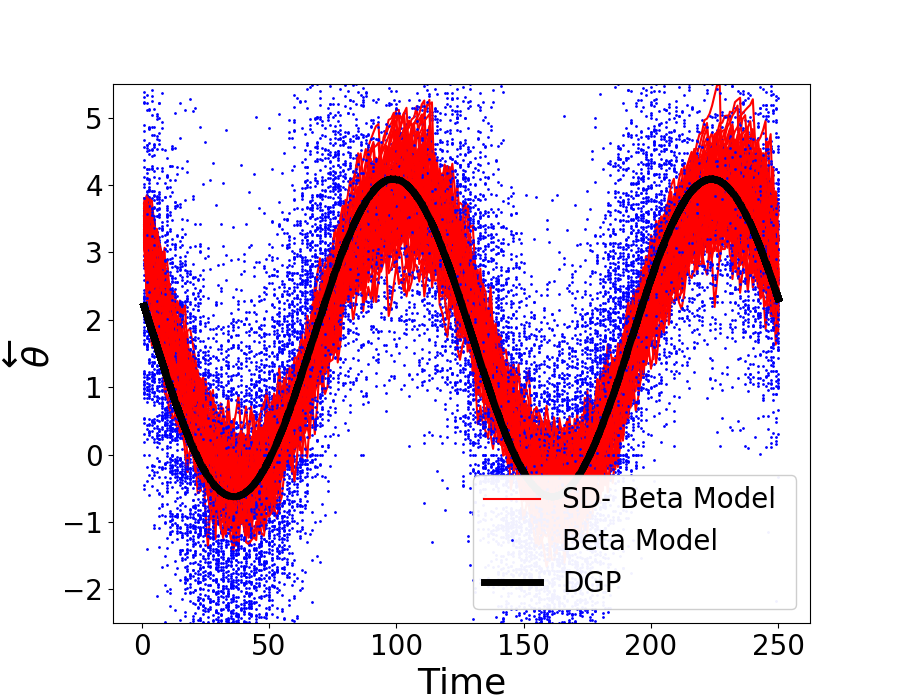} \\
	\includegraphics[width=0.45\linewidth 	]{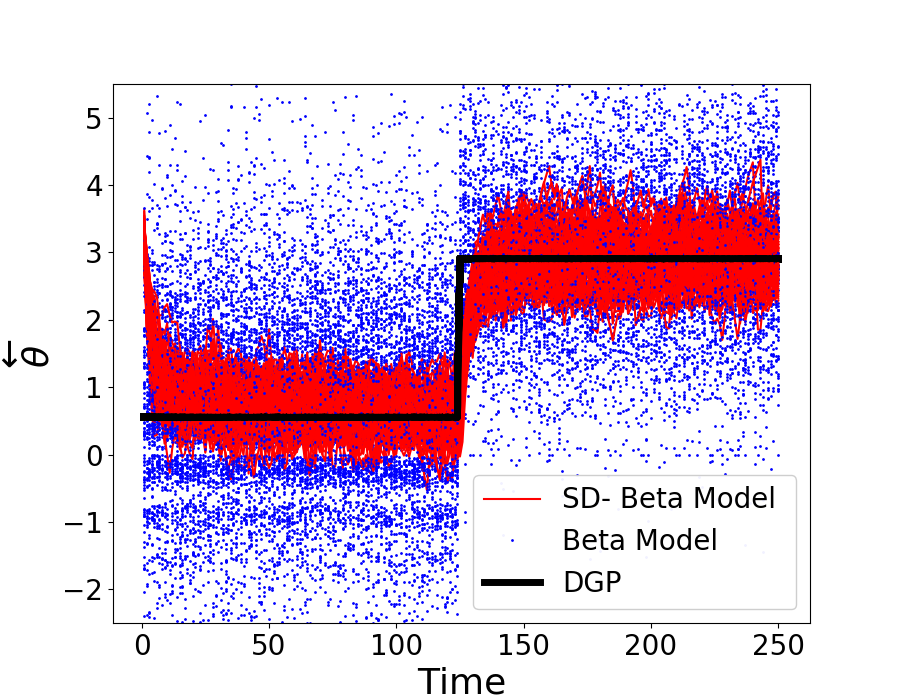}
	\includegraphics[width=0.45\linewidth 	]{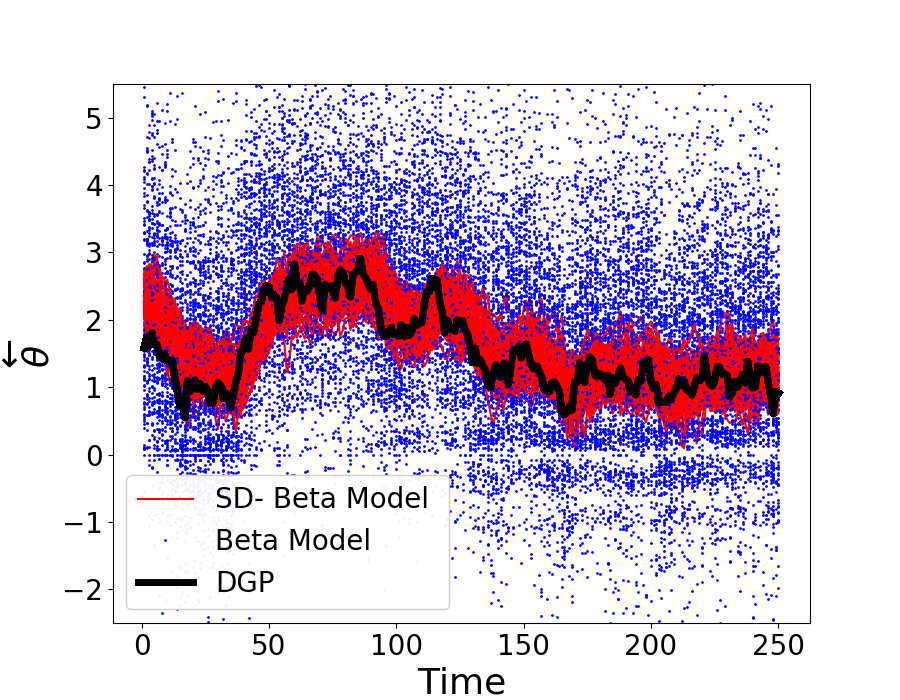} 
	\caption{Temporal evolution of one randomly selected parameter for the considered DGPs. The black line is the true path of the parameter of the DGP, the red ones are those filtered using the SD-beta model, and the blue dots correspond to the cross-sectional estimates of the beta model.}\label{fig:rmse_beta_model_fixed_N} 
\end{figure}

Figure \ref{fig:rmse_beta_model_fixed_N} shows the temporal evolution for one randomly chosen parameter of the beta model for all the DGPs, together with the paths filtered from the observations using the SD beta model and the sequence of cross-sectional static estimates. The score-driven filtering and cross-sectional estimation are repeated over $100$ simulated network sequences. As discussed in the main text, the paths filtered with the SD beta model are, on average, much more accurate than those recovered from a standard beta model. 

\subsection{SD-Beta Model for large N}\label{sec:appendix_SDBeta}
In one of the numerical simulations we presented in the main text, we consider networks of increasing size. Here, we present some additional details on how the DGPs are defined for networks of increasing size. Practically, we have to fix the vectors $\overline{\theta}_{0_s}$, $\overline{\theta}_{1_s}$, and $\overline{\theta}_{2_s}$ in a similar way, with the only difference being the numerical values for $D_m$ and $D_M$. Specifically, in the sparse case we keep for each $N$ $D_m = 10$ and $D_M = 40$. In the dense case, we set $D_M = 0.8 N$, i.e., both the maximum degree and the average degree increase. 

One peculiarity of the beta model is that the number of parameters, i.e., the length of the vectors $ \ibinpar$ and $ \obinpar  $, increases with the number of nodes. Consistently, when we use the score-driven extension described so far, the length of the vectors $w$, $\alpha$, and $\beta$ increases too. 

Recall that the numerical values of $\theta_0$, $\theta_1$, and $\theta_2$ are chosen to fix the values of average degrees over time and the amplitude of their fluctuations, as described in Appendix \ref{sec:appendix_num_details}. For each value of $N$, we choose them to guarantee heterogeneity in the degrees across nodes and significant fluctuation in time. 
Most importantly, we set a maximum degree attainable for a node and let it depend on $N$ in two distinct ways, each corresponding to a different density regime: one generating \textit{sparse} networks and the other \textit{dense} ones. It is crucial to notice that the asymptotic results of Ref.~\onlinecite{chatterjee2011random} are expected to hold only in the dense case.

In the first numerical experiment testing the SD-beta model as a misspecified filter, we estimated a total of $60$ parameters, $6$ static parameters for each one of the ten nodes, $3$ for the time-varying in-degree and $3$ for the time-varying out-degree. Here, we present the further parameter restriction mentioned in the main text that proved useful when the number of nodes increases. Specifically, we assume that the parameters $\alpha^{\text{out}}$ and $\beta^{\text{out}}$ are common to all out-degree tvps $\obinpar\et$. Similarly, all in-degree tvps $\ibinpar\et$ share the same $\alpha^{\text{in}}$ and $\beta^{\text{in}}$. The coefficients ${w}_s^{\text{in}}$ and  ${w}_s^{\text{out}}$ remain node specific. The resulting update rule is
\begin{eqnarray*}
	\ibinpar_s\etp &=  {w}_s^{\text{in}} + {\beta}^{\text{in}} 	\ibinpar_s\et +  {\alpha}^{\text{in}} \frac{\sum_{i} \left(Y_{is}\et - p_{is}\et\right)}{\sqrt{\sum_i p_{is}\et\tonde{1 - p_{is}\et }}}  \\
	\obinpar_s\etp &= {w}_s^{\text{out}} + {\beta}^{\text{out}} 	\obinpar_s\et +  {\alpha}^{\text{out}} \frac{\sum_{i} \left(Y_{si}\et - p_{si}\et\right)}{\sqrt{\sum_i p_{si}\et\tonde{1 - p_{si}\et }}}.
\end{eqnarray*}
\section*{References}

\begin{thebibliography}{96}%
\makeatletter
\providecommand \@ifxundefined [1]{%
 \@ifx{#1\undefined}
}%
\providecommand \@ifnum [1]{%
 \ifnum #1\expandafter \@firstoftwo
 \else \expandafter \@secondoftwo
 \fi
}%
\providecommand \@ifx [1]{%
 \ifx #1\expandafter \@firstoftwo
 \else \expandafter \@secondoftwo
 \fi
}%
\providecommand \natexlab [1]{#1}%
\providecommand \enquote  [1]{``#1''}%
\providecommand \bibnamefont  [1]{#1}%
\providecommand \bibfnamefont [1]{#1}%
\providecommand \citenamefont [1]{#1}%
\providecommand \href@noop [0]{\@secondoftwo}%
\providecommand \href [0]{\begingroup \@sanitize@url \@href}%
\providecommand \@href[1]{\@@startlink{#1}\@@href}%
\providecommand \@@href[1]{\endgroup#1\@@endlink}%
\providecommand \@sanitize@url [0]{\catcode `\\12\catcode `\$12\catcode `\&12\catcode `\#12\catcode `\^12\catcode `\_12\catcode `\%12\relax}%
\providecommand \@@startlink[1]{}%
\providecommand \@@endlink[0]{}%
\providecommand \url  [0]{\begingroup\@sanitize@url \@url }%
\providecommand \@url [1]{\endgroup\@href {#1}{\urlprefix }}%
\providecommand \urlprefix  [0]{URL }%
\providecommand \Eprint [0]{\href }%
\providecommand \doibase [0]{http://dx.doi.org/}%
\providecommand \selectlanguage [0]{\@gobble}%
\providecommand \bibinfo  [0]{\@secondoftwo}%
\providecommand \bibfield  [0]{\@secondoftwo}%
\providecommand \translation [1]{[#1]}%
\providecommand \BibitemOpen [0]{}%
\providecommand \bibitemStop [0]{}%
\providecommand \bibitemNoStop [0]{.\EOS\space}%
\providecommand \EOS [0]{\spacefactor3000\relax}%
\providecommand \BibitemShut  [1]{\csname bibitem#1\endcsname}%
\let\auto@bib@innerbib\@empty
\bibitem [{\citenamefont {Albert}\ and\ \citenamefont {Barab\'asi}(2002)}]{RevModPhys.74.47}%
  \BibitemOpen
  \bibfield  {author} {\bibinfo {author} {\bibfnamefont {R.}~\bibnamefont {Albert}}\ and\ \bibinfo {author} {\bibfnamefont {A.-L.}\ \bibnamefont {Barab\'asi}},\ }\bibfield  {title} {\enquote {\bibinfo {title} {Statistical mechanics of complex networks},}\ }\href {\doibase 10.1103/RevModPhys.74.47} {\bibfield  {journal} {\bibinfo  {journal} {Rev. Mod. Phys.}\ }\textbf {\bibinfo {volume} {74}},\ \bibinfo {pages} {47--97} (\bibinfo {year} {2002})}\BibitemShut {NoStop}%
\bibitem [{\citenamefont {Bullmore}\ and\ \citenamefont {Sporns}(2009)}]{bullmore2009complex}%
  \BibitemOpen
  \bibfield  {author} {\bibinfo {author} {\bibfnamefont {E.}~\bibnamefont {Bullmore}}\ and\ \bibinfo {author} {\bibfnamefont {O.}~\bibnamefont {Sporns}},\ }\bibfield  {title} {\enquote {\bibinfo {title} {Complex brain networks: graph theoretical analysis of structural and functional systems},}\ }\href@noop {} {\bibfield  {journal} {\bibinfo  {journal} {Nature Reviews Neuroscience}\ }\textbf {\bibinfo {volume} {10}},\ \bibinfo {pages} {186} (\bibinfo {year} {2009})}\BibitemShut {NoStop}%
\bibitem [{\citenamefont {Newman}(2010)}]{newman2010networks}%
  \BibitemOpen
  \bibfield  {author} {\bibinfo {author} {\bibfnamefont {M.}~\bibnamefont {Newman}},\ }\href@noop {} {\emph {\bibinfo {title} {Networks: an introduction}}}\ (\bibinfo  {publisher} {Oxford University Press},\ \bibinfo {year} {2010})\BibitemShut {NoStop}%
\bibitem [{\citenamefont {Easley}, \citenamefont {Kleinberg}\ \emph {et~al.}(2010)\citenamefont {Easley}, \citenamefont {Kleinberg} \emph {et~al.}}]{easley2010networks}%
  \BibitemOpen
  \bibfield  {author} {\bibinfo {author} {\bibfnamefont {D.}~\bibnamefont {Easley}}, \bibinfo {author} {\bibfnamefont {J.}~\bibnamefont {Kleinberg}},  \emph {et~al.},\ }\href@noop {} {\emph {\bibinfo {title} {Networks, crowds, and markets}}},\ Vol.~\bibinfo {volume} {8}\ (\bibinfo  {publisher} {Cambridge University Press Cambridge},\ \bibinfo {year} {2010})\BibitemShut {NoStop}%
\bibitem [{\citenamefont {Allen}\ and\ \citenamefont {Babus}(2009)}]{allen2009networks}%
  \BibitemOpen
  \bibfield  {author} {\bibinfo {author} {\bibfnamefont {F.}~\bibnamefont {Allen}}\ and\ \bibinfo {author} {\bibfnamefont {A.}~\bibnamefont {Babus}},\ }\bibfield  {title} {\enquote {\bibinfo {title} {Networks in finance},}\ }in\ \href@noop {} {\emph {\bibinfo {booktitle} {The network challenge: strategy, profit, and risk in an interlinked world}}},\ \bibinfo {editor} {edited by\ \bibinfo {editor} {\bibfnamefont {P.~R.}\ \bibnamefont {Kleindorfer}}\ and\ \bibinfo {editor} {\bibfnamefont {Y.~J.}\ \bibnamefont {Wind}}}\ (\bibinfo  {publisher} {Pearson Education},\ \bibinfo {year} {2009})\ Chap.~\bibinfo {chapter} {21}, pp.\ \bibinfo {pages} {367--379}\BibitemShut {NoStop}%
\bibitem [{\citenamefont {Holme}\ and\ \citenamefont {Saram{\"a}ki}(2012)}]{holme2012temporal}%
  \BibitemOpen
  \bibfield  {author} {\bibinfo {author} {\bibfnamefont {P.}~\bibnamefont {Holme}}\ and\ \bibinfo {author} {\bibfnamefont {J.}~\bibnamefont {Saram{\"a}ki}},\ }\bibfield  {title} {\enquote {\bibinfo {title} {Temporal networks},}\ }\href@noop {} {\bibfield  {journal} {\bibinfo  {journal} {Physics reports}\ }\textbf {\bibinfo {volume} {519}},\ \bibinfo {pages} {97--125} (\bibinfo {year} {2012})}\BibitemShut {NoStop}%
\bibitem [{\citenamefont {Craig}\ and\ \citenamefont {Von~Peter}(2014)}]{craig2014interbank}%
  \BibitemOpen
  \bibfield  {author} {\bibinfo {author} {\bibfnamefont {B.}~\bibnamefont {Craig}}\ and\ \bibinfo {author} {\bibfnamefont {G.}~\bibnamefont {Von~Peter}},\ }\bibfield  {title} {\enquote {\bibinfo {title} {Interbank tiering and money center banks},}\ }\href@noop {} {\bibfield  {journal} {\bibinfo  {journal} {Journal of Financial Intermediation}\ }\textbf {\bibinfo {volume} {23}},\ \bibinfo {pages} {322--347} (\bibinfo {year} {2014})}\BibitemShut {NoStop}%
\bibitem [{\citenamefont {Rossetti}\ and\ \citenamefont {Cazabet}(2018)}]{rossetti2018community}%
  \BibitemOpen
  \bibfield  {author} {\bibinfo {author} {\bibfnamefont {G.}~\bibnamefont {Rossetti}}\ and\ \bibinfo {author} {\bibfnamefont {R.}~\bibnamefont {Cazabet}},\ }\bibfield  {title} {\enquote {\bibinfo {title} {Community discovery in dynamic networks: a survey},}\ }\href@noop {} {\bibfield  {journal} {\bibinfo  {journal} {ACM Computing Surveys (CSUR)}\ }\textbf {\bibinfo {volume} {51}},\ \bibinfo {pages} {35} (\bibinfo {year} {2018})}\BibitemShut {NoStop}%
\bibitem [{\citenamefont {Kim}\ \emph {et~al.}(2018)\citenamefont {Kim}, \citenamefont {Lee}, \citenamefont {Xue}, \citenamefont {Niu} \emph {et~al.}}]{kim2018review}%
  \BibitemOpen
  \bibfield  {author} {\bibinfo {author} {\bibfnamefont {B.}~\bibnamefont {Kim}}, \bibinfo {author} {\bibfnamefont {K.~H.}\ \bibnamefont {Lee}}, \bibinfo {author} {\bibfnamefont {L.}~\bibnamefont {Xue}}, \bibinfo {author} {\bibfnamefont {X.}~\bibnamefont {Niu}},  \emph {et~al.},\ }\bibfield  {title} {\enquote {\bibinfo {title} {A review of dynamic network models with latent variables},}\ }\href@noop {} {\bibfield  {journal} {\bibinfo  {journal} {Statistics Surveys}\ }\textbf {\bibinfo {volume} {12}},\ \bibinfo {pages} {105--135} (\bibinfo {year} {2018})}\BibitemShut {NoStop}%
\bibitem [{\citenamefont {Robins}\ and\ \citenamefont {Pattison}(2001)}]{robins2001random}%
  \BibitemOpen
  \bibfield  {author} {\bibinfo {author} {\bibfnamefont {G.}~\bibnamefont {Robins}}\ and\ \bibinfo {author} {\bibfnamefont {P.}~\bibnamefont {Pattison}},\ }\bibfield  {title} {\enquote {\bibinfo {title} {Random graph models for temporal processes in social networks},}\ }\href@noop {} {\bibfield  {journal} {\bibinfo  {journal} {Journal of Mathematical Sociology}\ }\textbf {\bibinfo {volume} {25}},\ \bibinfo {pages} {5--41} (\bibinfo {year} {2001})}\BibitemShut {NoStop}%
\bibitem [{\citenamefont {Hanneke}\ \emph {et~al.}(2010)\citenamefont {Hanneke}, \citenamefont {Fu}, \citenamefont {Xing} \emph {et~al.}}]{hanneke2010discrete}%
  \BibitemOpen
  \bibfield  {author} {\bibinfo {author} {\bibfnamefont {S.}~\bibnamefont {Hanneke}}, \bibinfo {author} {\bibfnamefont {W.}~\bibnamefont {Fu}}, \bibinfo {author} {\bibfnamefont {E.~P.}\ \bibnamefont {Xing}},  \emph {et~al.},\ }\bibfield  {title} {\enquote {\bibinfo {title} {Discrete temporal models of social networks},}\ }\href@noop {} {\bibfield  {journal} {\bibinfo  {journal} {Electronic Journal of Statistics}\ }\textbf {\bibinfo {volume} {4}},\ \bibinfo {pages} {585--605} (\bibinfo {year} {2010})}\BibitemShut {NoStop}%
\bibitem [{\citenamefont {Cranmer}\ and\ \citenamefont {Desmarais}(2011)}]{cranmer2011inferential}%
  \BibitemOpen
  \bibfield  {author} {\bibinfo {author} {\bibfnamefont {S.~J.}\ \bibnamefont {Cranmer}}\ and\ \bibinfo {author} {\bibfnamefont {B.~A.}\ \bibnamefont {Desmarais}},\ }\bibfield  {title} {\enquote {\bibinfo {title} {Inferential network analysis with exponential random graph models},}\ }\href@noop {} {\bibfield  {journal} {\bibinfo  {journal} {Political Analysis}\ }\textbf {\bibinfo {volume} {19}},\ \bibinfo {pages} {66--86} (\bibinfo {year} {2011})}\BibitemShut {NoStop}%
\bibitem [{\citenamefont {Krivitsky}\ and\ \citenamefont {Handcock}(2014)}]{krivitsky2014separable}%
  \BibitemOpen
  \bibfield  {author} {\bibinfo {author} {\bibfnamefont {P.~N.}\ \bibnamefont {Krivitsky}}\ and\ \bibinfo {author} {\bibfnamefont {M.~S.}\ \bibnamefont {Handcock}},\ }\bibfield  {title} {\enquote {\bibinfo {title} {A separable model for dynamic networks},}\ }\href@noop {} {\bibfield  {journal} {\bibinfo  {journal} {Journal of the Royal Statistical Society: Series B (Statistical Methodology)}\ }\textbf {\bibinfo {volume} {76}},\ \bibinfo {pages} {29--46} (\bibinfo {year} {2014})}\BibitemShut {NoStop}%
\bibitem [{\citenamefont {Lee}, \citenamefont {Li},\ and\ \citenamefont {Wilson}(2020)}]{lee2020varying}%
  \BibitemOpen
  \bibfield  {author} {\bibinfo {author} {\bibfnamefont {J.}~\bibnamefont {Lee}}, \bibinfo {author} {\bibfnamefont {G.}~\bibnamefont {Li}}, \ and\ \bibinfo {author} {\bibfnamefont {J.~D.}\ \bibnamefont {Wilson}},\ }\bibfield  {title} {\enquote {\bibinfo {title} {Varying-coefficient models for dynamic networks},}\ }\href@noop {} {\bibfield  {journal} {\bibinfo  {journal} {Computational Statistics \& Data Analysis}\ }\textbf {\bibinfo {volume} {152}},\ \bibinfo {pages} {107052} (\bibinfo {year} {2020})}\BibitemShut {NoStop}%
\bibitem [{Note1()}]{Note1}%
  \BibitemOpen
  \bibinfo {note} {In time series jargon, it is a smoother and not a filter.}\BibitemShut {Stop}%
\bibitem [{\citenamefont {Mazzarisi}\ \emph {et~al.}(2020)\citenamefont {Mazzarisi}, \citenamefont {Barucca}, \citenamefont {Lillo},\ and\ \citenamefont {Tantari}}]{mazzarisi2020dynamic}%
  \BibitemOpen
  \bibfield  {author} {\bibinfo {author} {\bibfnamefont {P.}~\bibnamefont {Mazzarisi}}, \bibinfo {author} {\bibfnamefont {P.}~\bibnamefont {Barucca}}, \bibinfo {author} {\bibfnamefont {F.}~\bibnamefont {Lillo}}, \ and\ \bibinfo {author} {\bibfnamefont {D.}~\bibnamefont {Tantari}},\ }\bibfield  {title} {\enquote {\bibinfo {title} {A dynamic network model with persistent links and node-specific latent variables, with an application to the interbank market},}\ }\href@noop {} {\bibfield  {journal} {\bibinfo  {journal} {European Journal of Operational Research}\ }\textbf {\bibinfo {volume} {281}},\ \bibinfo {pages} {50--65} (\bibinfo {year} {2020})}\BibitemShut {NoStop}%
\bibitem [{\citenamefont {Cox}\ \emph {et~al.}(1981)\citenamefont {Cox}, \citenamefont {Gudmundsson}, \citenamefont {Lindgren}, \citenamefont {Bondesson}, \citenamefont {Harsaae}, \citenamefont {Laake}, \citenamefont {Juselius},\ and\ \citenamefont {Lauritzen}}]{cox1981statistical}%
  \BibitemOpen
  \bibfield  {author} {\bibinfo {author} {\bibfnamefont {D.~R.}\ \bibnamefont {Cox}}, \bibinfo {author} {\bibfnamefont {G.}~\bibnamefont {Gudmundsson}}, \bibinfo {author} {\bibfnamefont {G.}~\bibnamefont {Lindgren}}, \bibinfo {author} {\bibfnamefont {L.}~\bibnamefont {Bondesson}}, \bibinfo {author} {\bibfnamefont {E.}~\bibnamefont {Harsaae}}, \bibinfo {author} {\bibfnamefont {P.}~\bibnamefont {Laake}}, \bibinfo {author} {\bibfnamefont {K.}~\bibnamefont {Juselius}}, \ and\ \bibinfo {author} {\bibfnamefont {S.~L.}\ \bibnamefont {Lauritzen}},\ }\bibfield  {title} {\enquote {\bibinfo {title} {Statistical analysis of time series: Some recent developments},}\ }\href@noop {} {\bibfield  {journal} {\bibinfo  {journal} {Scandinavian Journal of Statistics}\ ,\ \bibinfo {pages} {93--115}} (\bibinfo {year} {1981})}\BibitemShut {NoStop}%
\bibitem [{\citenamefont {Snijders}(1996)}]{snijders1996stochastic}%
  \BibitemOpen
  \bibfield  {author} {\bibinfo {author} {\bibfnamefont {T.~A.}\ \bibnamefont {Snijders}},\ }\bibfield  {title} {\enquote {\bibinfo {title} {Stochastic actor-oriented models for network change},}\ }\href@noop {} {\bibfield  {journal} {\bibinfo  {journal} {Journal of Mathematical Sociology}\ }\textbf {\bibinfo {volume} {21}},\ \bibinfo {pages} {149--172} (\bibinfo {year} {1996})}\BibitemShut {NoStop}%
\bibitem [{\citenamefont {Butts}(2008)}]{butts20084}%
  \BibitemOpen
  \bibfield  {author} {\bibinfo {author} {\bibfnamefont {C.~T.}\ \bibnamefont {Butts}},\ }\bibfield  {title} {\enquote {\bibinfo {title} {A relational event framework for social action},}\ }\href@noop {} {\bibfield  {journal} {\bibinfo  {journal} {Sociological Methodology}\ }\textbf {\bibinfo {volume} {38}},\ \bibinfo {pages} {155--200} (\bibinfo {year} {2008})}\BibitemShut {NoStop}%
\bibitem [{\citenamefont {Kolaczyk}(2009)}]{Kolaczyk:2009:SAN:1593430}%
  \BibitemOpen
  \bibfield  {author} {\bibinfo {author} {\bibfnamefont {E.~D.}\ \bibnamefont {Kolaczyk}},\ }\href@noop {} {\emph {\bibinfo {title} {Statistical Analysis of Network Data: Methods and Models}}},\ \bibinfo {edition} {1st}\ ed.\ (\bibinfo  {publisher} {Springer Publishing Company, Incorporated},\ \bibinfo {year} {2009})\BibitemShut {NoStop}%
\bibitem [{\citenamefont {Barndorff-Nielsen}(2014)}]{barndorff2014information}%
  \BibitemOpen
  \bibfield  {author} {\bibinfo {author} {\bibfnamefont {O.}~\bibnamefont {Barndorff-Nielsen}},\ }\href@noop {} {\emph {\bibinfo {title} {Information and exponential families in statistical theory}}}\ (\bibinfo  {publisher} {John Wiley \& Sons},\ \bibinfo {year} {2014})\BibitemShut {NoStop}%
\bibitem [{\citenamefont {Schweinberger}\ \emph {et~al.}(2020)\citenamefont {Schweinberger}, \citenamefont {Krivitsky}, \citenamefont {Butts},\ and\ \citenamefont {Stewart}}]{schweinbergerexponential}%
  \BibitemOpen
  \bibfield  {author} {\bibinfo {author} {\bibfnamefont {M.}~\bibnamefont {Schweinberger}}, \bibinfo {author} {\bibfnamefont {P.~N.}\ \bibnamefont {Krivitsky}}, \bibinfo {author} {\bibfnamefont {C.~T.}\ \bibnamefont {Butts}}, \ and\ \bibinfo {author} {\bibfnamefont {J.~R.}\ \bibnamefont {Stewart}},\ }\bibfield  {title} {\enquote {\bibinfo {title} {Exponential-family models of random graphs: Inference in finite, super and infinite-population scenarios},}\ }\href@noop {} {\bibfield  {journal} {\bibinfo  {journal} {Statistical Science}\ }\textbf {\bibinfo {volume} {35}},\ \bibinfo {pages} {627--662} (\bibinfo {year} {2020})}\BibitemShut {NoStop}%
\bibitem [{\citenamefont {Shannon}(2001)}]{Shannon:2001:MTC:584091.584093}%
  \BibitemOpen
  \bibfield  {author} {\bibinfo {author} {\bibfnamefont {C.~E.}\ \bibnamefont {Shannon}},\ }\bibfield  {title} {\enquote {\bibinfo {title} {A mathematical theory of communication},}\ }\href {\doibase 10.1145/584091.584093} {\bibfield  {journal} {\bibinfo  {journal} {SIGMOBILE Mob. Comput. Commun. Rev.}\ }\textbf {\bibinfo {volume} {5}},\ \bibinfo {pages} {3--55} (\bibinfo {year} {2001})}\BibitemShut {NoStop}%
\bibitem [{\citenamefont {Jaynes}(1957)}]{PhysRev.106.620}%
  \BibitemOpen
  \bibfield  {author} {\bibinfo {author} {\bibfnamefont {E.}~\bibnamefont {Jaynes}},\ }\bibfield  {title} {\enquote {\bibinfo {title} {Information theory and statistical mechanics},}\ }\href {\doibase 10.1103/PhysRev.106.620} {\bibfield  {journal} {\bibinfo  {journal} {Phys. Rev.}\ }\textbf {\bibinfo {volume} {106}},\ \bibinfo {pages} {620--630} (\bibinfo {year} {1957})}\BibitemShut {NoStop}%
\bibitem [{\citenamefont {Park}\ and\ \citenamefont {Newman}(2004)}]{PhysRevE.70.066117}%
  \BibitemOpen
  \bibfield  {author} {\bibinfo {author} {\bibfnamefont {J.}~\bibnamefont {Park}}\ and\ \bibinfo {author} {\bibfnamefont {M.~E.~J.}\ \bibnamefont {Newman}},\ }\bibfield  {title} {\enquote {\bibinfo {title} {Statistical mechanics of networks},}\ }\href {\doibase 10.1103/PhysRevE.70.066117} {\bibfield  {journal} {\bibinfo  {journal} {Phys. Rev. E}\ }\textbf {\bibinfo {volume} {70}},\ \bibinfo {pages} {066117} (\bibinfo {year} {2004})}\BibitemShut {NoStop}%
\bibitem [{\citenamefont {Garlaschelli}\ and\ \citenamefont {Loffredo}(2008)}]{PhysRevE.78.015101}%
  \BibitemOpen
  \bibfield  {author} {\bibinfo {author} {\bibfnamefont {D.}~\bibnamefont {Garlaschelli}}\ and\ \bibinfo {author} {\bibfnamefont {M.~I.}\ \bibnamefont {Loffredo}},\ }\bibfield  {title} {\enquote {\bibinfo {title} {Maximum likelihood: Extracting unbiased information from complex networks},}\ }\href {\doibase 10.1103/PhysRevE.78.015101} {\bibfield  {journal} {\bibinfo  {journal} {Phys. Rev. E}\ }\textbf {\bibinfo {volume} {78}},\ \bibinfo {pages} {015101} (\bibinfo {year} {2008})}\BibitemShut {NoStop}%
\bibitem [{\citenamefont {Chatterjee}\ \emph {et~al.}(2011)\citenamefont {Chatterjee}, \citenamefont {Diaconis}, \citenamefont {Sly} \emph {et~al.}}]{chatterjee2011random}%
  \BibitemOpen
  \bibfield  {author} {\bibinfo {author} {\bibfnamefont {S.}~\bibnamefont {Chatterjee}}, \bibinfo {author} {\bibfnamefont {P.}~\bibnamefont {Diaconis}}, \bibinfo {author} {\bibfnamefont {A.}~\bibnamefont {Sly}},  \emph {et~al.},\ }\bibfield  {title} {\enquote {\bibinfo {title} {Random graphs with a given degree sequence},}\ }\href@noop {} {\bibfield  {journal} {\bibinfo  {journal} {The Annals of Applied Probability}\ }\textbf {\bibinfo {volume} {21}},\ \bibinfo {pages} {1400--1435} (\bibinfo {year} {2011})}\BibitemShut {NoStop}%
\bibitem [{\citenamefont {Zermelo}(1929)}]{zermelo1929berechnung}%
  \BibitemOpen
  \bibfield  {author} {\bibinfo {author} {\bibfnamefont {E.}~\bibnamefont {Zermelo}},\ }\bibfield  {title} {\enquote {\bibinfo {title} {Die berechnung der turnier-ergebnisse als ein maximumproblem der wahrscheinlichkeitsrechnung},}\ }\href@noop {} {\bibfield  {journal} {\bibinfo  {journal} {Mathematische Zeitschrift}\ }\textbf {\bibinfo {volume} {29}},\ \bibinfo {pages} {436--460} (\bibinfo {year} {1929})}\BibitemShut {NoStop}%
\bibitem [{\citenamefont {Holland}\ and\ \citenamefont {Leinhardt}(1981)}]{Holland81anexponential}%
  \BibitemOpen
  \bibfield  {author} {\bibinfo {author} {\bibfnamefont {P.~W.}\ \bibnamefont {Holland}}\ and\ \bibinfo {author} {\bibfnamefont {S.}~\bibnamefont {Leinhardt}},\ }\bibfield  {title} {\enquote {\bibinfo {title} {An exponential family of probability distributions for directed graphs},}\ }\href@noop {} {\bibfield  {journal} {\bibinfo  {journal} {Journal of the American Statistical Association}\ }\textbf {\bibinfo {volume} {76}},\ \bibinfo {pages} {33--50} (\bibinfo {year} {1981})}\BibitemShut {NoStop}%
\bibitem [{\citenamefont {Caldarelli}\ \emph {et~al.}(2002)\citenamefont {Caldarelli}, \citenamefont {Capocci}, \citenamefont {De~Los~Rios},\ and\ \citenamefont {Mu\~noz}}]{PhysRevLett.89.258702}%
  \BibitemOpen
  \bibfield  {author} {\bibinfo {author} {\bibfnamefont {G.}~\bibnamefont {Caldarelli}}, \bibinfo {author} {\bibfnamefont {A.}~\bibnamefont {Capocci}}, \bibinfo {author} {\bibfnamefont {P.}~\bibnamefont {De~Los~Rios}}, \ and\ \bibinfo {author} {\bibfnamefont {M.~A.}\ \bibnamefont {Mu\~noz}},\ }\bibfield  {title} {\enquote {\bibinfo {title} {Scale-free networks from varying vertex intrinsic fitness},}\ }\href {\doibase 10.1103/PhysRevLett.89.258702} {\bibfield  {journal} {\bibinfo  {journal} {Phys. Rev. Lett.}\ }\textbf {\bibinfo {volume} {89}},\ \bibinfo {pages} {258702} (\bibinfo {year} {2002})}\BibitemShut {NoStop}%
\bibitem [{\citenamefont {Snijders}\ \emph {et~al.}(2006)\citenamefont {Snijders}, \citenamefont {Pattison}, \citenamefont {Robins},\ and\ \citenamefont {Handcock}}]{snijders2006new}%
  \BibitemOpen
  \bibfield  {author} {\bibinfo {author} {\bibfnamefont {T.~A.}\ \bibnamefont {Snijders}}, \bibinfo {author} {\bibfnamefont {P.~E.}\ \bibnamefont {Pattison}}, \bibinfo {author} {\bibfnamefont {G.~L.}\ \bibnamefont {Robins}}, \ and\ \bibinfo {author} {\bibfnamefont {M.~S.}\ \bibnamefont {Handcock}},\ }\bibfield  {title} {\enquote {\bibinfo {title} {New specifications for exponential random graph models},}\ }\href@noop {} {\bibfield  {journal} {\bibinfo  {journal} {Sociological Methodology}\ }\textbf {\bibinfo {volume} {36}},\ \bibinfo {pages} {99--153} (\bibinfo {year} {2006})}\BibitemShut {NoStop}%
\bibitem [{\citenamefont {Robins}\ \emph {et~al.}(2007)\citenamefont {Robins}, \citenamefont {Snijders}, \citenamefont {Wang}, \citenamefont {Handcock},\ and\ \citenamefont {Pattison}}]{robins2007recent}%
  \BibitemOpen
  \bibfield  {author} {\bibinfo {author} {\bibfnamefont {G.}~\bibnamefont {Robins}}, \bibinfo {author} {\bibfnamefont {T.}~\bibnamefont {Snijders}}, \bibinfo {author} {\bibfnamefont {P.}~\bibnamefont {Wang}}, \bibinfo {author} {\bibfnamefont {M.}~\bibnamefont {Handcock}}, \ and\ \bibinfo {author} {\bibfnamefont {P.}~\bibnamefont {Pattison}},\ }\bibfield  {title} {\enquote {\bibinfo {title} {Recent developments in exponential random graph (p*) models for social networks},}\ }\href@noop {} {\bibfield  {journal} {\bibinfo  {journal} {Social Networks}\ }\textbf {\bibinfo {volume} {29}},\ \bibinfo {pages} {192--215} (\bibinfo {year} {2007})}\BibitemShut {NoStop}%
\bibitem [{\citenamefont {Hunter}\ and\ \citenamefont {Handcock}(2006)}]{hunter2006inference}%
  \BibitemOpen
  \bibfield  {author} {\bibinfo {author} {\bibfnamefont {D.~R.}\ \bibnamefont {Hunter}}\ and\ \bibinfo {author} {\bibfnamefont {M.~S.}\ \bibnamefont {Handcock}},\ }\bibfield  {title} {\enquote {\bibinfo {title} {Inference in curved exponential family models for networks},}\ }\href@noop {} {\bibfield  {journal} {\bibinfo  {journal} {Journal of Computational and Graphical Statistics}\ }\textbf {\bibinfo {volume} {15}},\ \bibinfo {pages} {565--583} (\bibinfo {year} {2006})}\BibitemShut {NoStop}%
\bibitem [{\citenamefont {Creal}, \citenamefont {Koopman},\ and\ \citenamefont {Lucas}(2013)}]{creal2013generalized}%
  \BibitemOpen
  \bibfield  {author} {\bibinfo {author} {\bibfnamefont {D.}~\bibnamefont {Creal}}, \bibinfo {author} {\bibfnamefont {S.~J.}\ \bibnamefont {Koopman}}, \ and\ \bibinfo {author} {\bibfnamefont {A.}~\bibnamefont {Lucas}},\ }\bibfield  {title} {\enquote {\bibinfo {title} {Generalized autoregressive score models with applications},}\ }\href@noop {} {\bibfield  {journal} {\bibinfo  {journal} {Journal of Applied Econometrics}\ }\textbf {\bibinfo {volume} {28}},\ \bibinfo {pages} {777--795} (\bibinfo {year} {2013})}\BibitemShut {NoStop}%
\bibitem [{\citenamefont {Harvey}(2013)}]{harvey2013dynamic}%
  \BibitemOpen
  \bibfield  {author} {\bibinfo {author} {\bibfnamefont {A.~C.}\ \bibnamefont {Harvey}},\ }\href@noop {} {\emph {\bibinfo {title} {Dynamic models for volatility and heavy tails: with applications to financial and economic time series}}},\ Vol.~\bibinfo {volume} {52}\ (\bibinfo  {publisher} {Cambridge University Press},\ \bibinfo {year} {2013})\BibitemShut {NoStop}%
\bibitem [{Note2()}]{Note2}%
  \BibitemOpen
  \bibinfo {note} {\protect \texttt {http://www.gasmodel.com/index.htm} for the updated collection of papers dealing with GAS models.}\BibitemShut {Stop}%
\bibitem [{\citenamefont {Blasques}, \citenamefont {Koopman},\ and\ \citenamefont {Lucas}(2015)}]{blasques2015information}%
  \BibitemOpen
  \bibfield  {author} {\bibinfo {author} {\bibfnamefont {F.}~\bibnamefont {Blasques}}, \bibinfo {author} {\bibfnamefont {S.~J.}\ \bibnamefont {Koopman}}, \ and\ \bibinfo {author} {\bibfnamefont {A.}~\bibnamefont {Lucas}},\ }\bibfield  {title} {\enquote {\bibinfo {title} {Information-theoretic optimality of observation-driven time series models for continuous responses},}\ }\href@noop {} {\bibfield  {journal} {\bibinfo  {journal} {Biometrika}\ }\textbf {\bibinfo {volume} {102}},\ \bibinfo {pages} {325--343} (\bibinfo {year} {2015})}\BibitemShut {NoStop}%
\bibitem [{\citenamefont {Bollerslev}(1986)}]{BOLLERSLEV1986307}%
  \BibitemOpen
  \bibfield  {author} {\bibinfo {author} {\bibfnamefont {T.}~\bibnamefont {Bollerslev}},\ }\bibfield  {title} {\enquote {\bibinfo {title} {Generalized autoregressive conditional heteroskedasticity},}\ }\href {\doibase https://doi.org/10.1016/0304-4076(86)90063-1} {\bibfield  {journal} {\bibinfo  {journal} {Journal of Econometrics}\ }\textbf {\bibinfo {volume} {31}},\ \bibinfo {pages} {307 -- 327} (\bibinfo {year} {1986})}\BibitemShut {NoStop}%
\bibitem [{\citenamefont {Nelson}(1991)}]{nelson1991conditional}%
  \BibitemOpen
  \bibfield  {author} {\bibinfo {author} {\bibfnamefont {D.~B.}\ \bibnamefont {Nelson}},\ }\bibfield  {title} {\enquote {\bibinfo {title} {Conditional heteroskedasticity in asset returns: A new approach},}\ }\href@noop {} {\bibfield  {journal} {\bibinfo  {journal} {Econometrica: Journal of the Econometric Society}\ ,\ \bibinfo {pages} {347--370}} (\bibinfo {year} {1991})}\BibitemShut {NoStop}%
\bibitem [{\citenamefont {Engle}\ and\ \citenamefont {Russell}(1998)}]{engle1998autoregressive}%
  \BibitemOpen
  \bibfield  {author} {\bibinfo {author} {\bibfnamefont {R.~F.}\ \bibnamefont {Engle}}\ and\ \bibinfo {author} {\bibfnamefont {J.~R.}\ \bibnamefont {Russell}},\ }\bibfield  {title} {\enquote {\bibinfo {title} {Autoregressive conditional duration: a new model for irregularly spaced transaction data},}\ }\href@noop {} {\bibfield  {journal} {\bibinfo  {journal} {Econometrica}\ ,\ \bibinfo {pages} {1127--1162}} (\bibinfo {year} {1998})}\BibitemShut {NoStop}%
\bibitem [{\citenamefont {Engle}(2002)}]{engle2002new}%
  \BibitemOpen
  \bibfield  {author} {\bibinfo {author} {\bibfnamefont {R.}~\bibnamefont {Engle}},\ }\bibfield  {title} {\enquote {\bibinfo {title} {New frontiers for {ARCH} models},}\ }\href@noop {} {\bibfield  {journal} {\bibinfo  {journal} {Journal of Applied Econometrics}\ }\textbf {\bibinfo {volume} {17}},\ \bibinfo {pages} {425--446} (\bibinfo {year} {2002})}\BibitemShut {NoStop}%
\bibitem [{\citenamefont {Goodreau}(2007)}]{goodreau2007advances}%
  \BibitemOpen
  \bibfield  {author} {\bibinfo {author} {\bibfnamefont {S.~M.}\ \bibnamefont {Goodreau}},\ }\bibfield  {title} {\enquote {\bibinfo {title} {Advances in exponential random graph (p*) models applied to a large social network},}\ }\href@noop {} {\bibfield  {journal} {\bibinfo  {journal} {Social Networks}\ }\textbf {\bibinfo {volume} {29}},\ \bibinfo {pages} {231--248} (\bibinfo {year} {2007})}\BibitemShut {NoStop}%
\bibitem [{\citenamefont {Hunter}, \citenamefont {Goodreau},\ and\ \citenamefont {Handcock}(2008)}]{hunter2008goodness}%
  \BibitemOpen
  \bibfield  {author} {\bibinfo {author} {\bibfnamefont {D.~R.}\ \bibnamefont {Hunter}}, \bibinfo {author} {\bibfnamefont {S.~M.}\ \bibnamefont {Goodreau}}, \ and\ \bibinfo {author} {\bibfnamefont {M.~S.}\ \bibnamefont {Handcock}},\ }\bibfield  {title} {\enquote {\bibinfo {title} {Goodness of fit of social network models},}\ }\href@noop {} {\bibfield  {journal} {\bibinfo  {journal} {Journal of the American Statistical Association}\ }\textbf {\bibinfo {volume} {103}},\ \bibinfo {pages} {248--258} (\bibinfo {year} {2008})}\BibitemShut {NoStop}%
\bibitem [{\citenamefont {Shore}\ and\ \citenamefont {Lubin}(2015)}]{SHORE201516}%
  \BibitemOpen
  \bibfield  {author} {\bibinfo {author} {\bibfnamefont {J.}~\bibnamefont {Shore}}\ and\ \bibinfo {author} {\bibfnamefont {B.}~\bibnamefont {Lubin}},\ }\bibfield  {title} {\enquote {\bibinfo {title} {Spectral goodness of fit for network models},}\ }\href {\doibase https://doi.org/10.1016/j.socnet.2015.04.004} {\bibfield  {journal} {\bibinfo  {journal} {Social Networks}\ }\textbf {\bibinfo {volume} {43}},\ \bibinfo {pages} {16 -- 27} (\bibinfo {year} {2015})}\BibitemShut {NoStop}%
\bibitem [{\citenamefont {Calvori}\ \emph {et~al.}(2017)\citenamefont {Calvori}, \citenamefont {Creal}, \citenamefont {Koopman},\ and\ \citenamefont {Lucas}}]{calvori2017testing}%
  \BibitemOpen
  \bibfield  {author} {\bibinfo {author} {\bibfnamefont {F.}~\bibnamefont {Calvori}}, \bibinfo {author} {\bibfnamefont {D.}~\bibnamefont {Creal}}, \bibinfo {author} {\bibfnamefont {S.~J.}\ \bibnamefont {Koopman}}, \ and\ \bibinfo {author} {\bibfnamefont {A.}~\bibnamefont {Lucas}},\ }\bibfield  {title} {\enquote {\bibinfo {title} {Testing for parameter instability across different modeling frameworks},}\ }\href@noop {} {\bibfield  {journal} {\bibinfo  {journal} {Journal of Financial Econometrics}\ }\textbf {\bibinfo {volume} {15}},\ \bibinfo {pages} {223--246} (\bibinfo {year} {2017})}\BibitemShut {NoStop}%
\bibitem [{\citenamefont {Nelson}(1996)}]{Nelson96}%
  \BibitemOpen
  \bibfield  {author} {\bibinfo {author} {\bibfnamefont {D.~B.}\ \bibnamefont {Nelson}},\ }\bibfield  {title} {\enquote {\bibinfo {title} {Asymptotically optimal smoothing with {ARCH} models},}\ }\href@noop {} {\bibfield  {journal} {\bibinfo  {journal} {Econometrica}\ }\textbf {\bibinfo {volume} {64}},\ \bibinfo {pages} {561–573} (\bibinfo {year} {1996})}\BibitemShut {NoStop}%
\bibitem [{Note3()}]{Note3}%
  \BibitemOpen
  \bibinfo {note} {In the following, the notation with a bar refers to the true parameters used in the DGP.}\BibitemShut {Stop}%
\bibitem [{Note4()}]{Note4}%
  \BibitemOpen
  \bibinfo {note} {For practical applications, it is very convenient that, for a large number of network functions, an efficient implementation to compute change statistics is made available in the R package \protect \textit {ergm}~\cite {hunter2008ergm}.}\BibitemShut {Stop}%
\bibitem [{\citenamefont {Hunter}\ \emph {et~al.}(2008)\citenamefont {Hunter}, \citenamefont {Handcock}, \citenamefont {Butts}, \citenamefont {Goodreau},\ and\ \citenamefont {Morris}}]{hunter2008ergm}%
  \BibitemOpen
  \bibfield  {author} {\bibinfo {author} {\bibfnamefont {D.~R.}\ \bibnamefont {Hunter}}, \bibinfo {author} {\bibfnamefont {M.~S.}\ \bibnamefont {Handcock}}, \bibinfo {author} {\bibfnamefont {C.~T.}\ \bibnamefont {Butts}}, \bibinfo {author} {\bibfnamefont {S.~M.}\ \bibnamefont {Goodreau}}, \ and\ \bibinfo {author} {\bibfnamefont {M.}~\bibnamefont {Morris}},\ }\bibfield  {title} {\enquote {\bibinfo {title} {ergm: A package to fit, simulate and diagnose exponential-family models for networks},}\ }\href@noop {} {\bibfield  {journal} {\bibinfo  {journal} {Journal of Statistical Software}\ }\textbf {\bibinfo {volume} {24}} (\bibinfo {year} {2008})}\BibitemShut {NoStop}%
\bibitem [{\citenamefont {Snijders}(2002)}]{snijders2002markov}%
  \BibitemOpen
  \bibfield  {author} {\bibinfo {author} {\bibfnamefont {T.~A.}\ \bibnamefont {Snijders}},\ }\bibfield  {title} {\enquote {\bibinfo {title} {Markov {C}hain {M}onte {C}arlo estimation of exponential random graph models},}\ }\href@noop {} {\bibfield  {journal} {\bibinfo  {journal} {Journal of Social Structure}\ }\textbf {\bibinfo {volume} {3}},\ \bibinfo {pages} {1--40} (\bibinfo {year} {2002})}\BibitemShut {NoStop}%
\bibitem [{\citenamefont {Buccheri}\ \emph {et~al.}(2018)\citenamefont {Buccheri}, \citenamefont {Bormetti}, \citenamefont {Corsi},\ and\ \citenamefont {Lillo}}]{buccheri_etal2018smoother}%
  \BibitemOpen
  \bibfield  {author} {\bibinfo {author} {\bibfnamefont {G.}~\bibnamefont {Buccheri}}, \bibinfo {author} {\bibfnamefont {G.}~\bibnamefont {Bormetti}}, \bibinfo {author} {\bibfnamefont {F.}~\bibnamefont {Corsi}}, \ and\ \bibinfo {author} {\bibfnamefont {F.}~\bibnamefont {Lillo}},\ }\bibfield  {title} {\enquote {\bibinfo {title} {Filtering and smoothing with score-driven models},}\ }\href@noop {} {\bibfield  {journal} {\bibinfo  {journal} {Available at SSRN: https://ssrn.com/abstract=3139666}\ } (\bibinfo {year} {2018})}\BibitemShut {NoStop}%
\bibitem [{Note5()}]{Note5}%
  \BibitemOpen
  \bibinfo {note} {We found the conclusions of this section to hold also in a sparse network density regime.}\BibitemShut {Stop}%
\bibitem [{\citenamefont {Iori}\ \emph {et~al.}(2008)\citenamefont {Iori}, \citenamefont {De~Masi}, \citenamefont {Precup}, \citenamefont {Gabbi},\ and\ \citenamefont {Caldarelli}}]{iori2008network}%
  \BibitemOpen
  \bibfield  {author} {\bibinfo {author} {\bibfnamefont {G.}~\bibnamefont {Iori}}, \bibinfo {author} {\bibfnamefont {G.}~\bibnamefont {De~Masi}}, \bibinfo {author} {\bibfnamefont {O.~V.}\ \bibnamefont {Precup}}, \bibinfo {author} {\bibfnamefont {G.}~\bibnamefont {Gabbi}}, \ and\ \bibinfo {author} {\bibfnamefont {G.}~\bibnamefont {Caldarelli}},\ }\bibfield  {title} {\enquote {\bibinfo {title} {A network analysis of the italian overnight money market},}\ }\href@noop {} {\bibfield  {journal} {\bibinfo  {journal} {J. Econ. Dyn. Control}\ }\textbf {\bibinfo {volume} {32}},\ \bibinfo {pages} {259--278} (\bibinfo {year} {2008})}\BibitemShut {NoStop}%
\bibitem [{\citenamefont {Finger}, \citenamefont {Fricke},\ and\ \citenamefont {Lux}(2013)}]{finger2013network}%
  \BibitemOpen
  \bibfield  {author} {\bibinfo {author} {\bibfnamefont {K.}~\bibnamefont {Finger}}, \bibinfo {author} {\bibfnamefont {D.}~\bibnamefont {Fricke}}, \ and\ \bibinfo {author} {\bibfnamefont {T.}~\bibnamefont {Lux}},\ }\bibfield  {title} {\enquote {\bibinfo {title} {Network analysis of the e-mid overnight money market: the informational value of different aggregation levels for intrinsic dynamic processes},}\ }\href@noop {} {\bibfield  {journal} {\bibinfo  {journal} {Computational Management Science}\ }\textbf {\bibinfo {volume} {10}},\ \bibinfo {pages} {187--211} (\bibinfo {year} {2013})}\BibitemShut {NoStop}%
\bibitem [{\citenamefont {Barucca}\ and\ \citenamefont {Lillo}(2018)}]{barucca2018organization}%
  \BibitemOpen
  \bibfield  {author} {\bibinfo {author} {\bibfnamefont {P.}~\bibnamefont {Barucca}}\ and\ \bibinfo {author} {\bibfnamefont {F.}~\bibnamefont {Lillo}},\ }\bibfield  {title} {\enquote {\bibinfo {title} {The organization of the interbank network and how {ECB} unconventional measures affected the e-{MID} overnight market},}\ }\href@noop {} {\bibfield  {journal} {\bibinfo  {journal} {Computational Management Science}\ }\textbf {\bibinfo {volume} {15}},\ \bibinfo {pages} {33--53} (\bibinfo {year} {2018})}\BibitemShut {NoStop}%
\bibitem [{Note6()}]{Note6}%
  \BibitemOpen
  \bibinfo {note} {It is worth stressing that the results become stable after $20$ simulations.}\BibitemShut {Stop}%
\bibitem [{Note7()}]{Note7}%
  \BibitemOpen
  \bibinfo {note} {In all the results on link forecasting -- one- or multi-step-ahead -- we excluded the links that are always zero, i.e., they never appear in the train and test samples. The reason is that those are extremely easy to predict, and keeping them would give an unrealistically optimistic picture of the predictability of links in the data set. Notably, the ranking of the methods remains unaltered when we keep all links for performance evaluation.}\BibitemShut {Stop}%
\bibitem [{\citenamefont {Fowler}(2006)}]{fowler2006connecting}%
  \BibitemOpen
  \bibfield  {author} {\bibinfo {author} {\bibfnamefont {J.~H.}\ \bibnamefont {Fowler}},\ }\bibfield  {title} {\enquote {\bibinfo {title} {Connecting the congress: A study of cosponsorship networks},}\ }\href@noop {} {\bibfield  {journal} {\bibinfo  {journal} {Political Analysis}\ }\textbf {\bibinfo {volume} {14}},\ \bibinfo {pages} {456--487} (\bibinfo {year} {2006})}\BibitemShut {NoStop}%
\bibitem [{\citenamefont {Faust}\ and\ \citenamefont {Skvoretz}(2002)}]{faust2002comparing}%
  \BibitemOpen
  \bibfield  {author} {\bibinfo {author} {\bibfnamefont {K.}~\bibnamefont {Faust}}\ and\ \bibinfo {author} {\bibfnamefont {J.}~\bibnamefont {Skvoretz}},\ }\bibfield  {title} {\enquote {\bibinfo {title} {Comparing networks across space and time, size and species},}\ }\href@noop {} {\bibfield  {journal} {\bibinfo  {journal} {Sociological Methodology}\ }\textbf {\bibinfo {volume} {32}},\ \bibinfo {pages} {267--299} (\bibinfo {year} {2002})}\BibitemShut {NoStop}%
\bibitem [{\citenamefont {Zhang}\ \emph {et~al.}(2008)\citenamefont {Zhang}, \citenamefont {Friend}, \citenamefont {Traud}, \citenamefont {Porter}, \citenamefont {Fowler},\ and\ \citenamefont {Mucha}}]{zhang2008community}%
  \BibitemOpen
  \bibfield  {author} {\bibinfo {author} {\bibfnamefont {Y.}~\bibnamefont {Zhang}}, \bibinfo {author} {\bibfnamefont {A.~J.}\ \bibnamefont {Friend}}, \bibinfo {author} {\bibfnamefont {A.~L.}\ \bibnamefont {Traud}}, \bibinfo {author} {\bibfnamefont {M.~A.}\ \bibnamefont {Porter}}, \bibinfo {author} {\bibfnamefont {J.~H.}\ \bibnamefont {Fowler}}, \ and\ \bibinfo {author} {\bibfnamefont {P.~J.}\ \bibnamefont {Mucha}},\ }\bibfield  {title} {\enquote {\bibinfo {title} {Community structure in congressional cosponsorship networks},}\ }\href@noop {} {\bibfield  {journal} {\bibinfo  {journal} {Physica A: Statistical Mechanics and its Applications}\ }\textbf {\bibinfo {volume} {387}},\ \bibinfo {pages} {1705--1712} (\bibinfo {year} {2008})}\BibitemShut {NoStop}%
\bibitem [{\citenamefont {Moody}\ and\ \citenamefont {Mucha}(2013)}]{moody2013portrait}%
  \BibitemOpen
  \bibfield  {author} {\bibinfo {author} {\bibfnamefont {J.}~\bibnamefont {Moody}}\ and\ \bibinfo {author} {\bibfnamefont {P.~J.}\ \bibnamefont {Mucha}},\ }\bibfield  {title} {\enquote {\bibinfo {title} {Portrait of political party polarization},}\ }\href@noop {} {\bibfield  {journal} {\bibinfo  {journal} {Network Science}\ }\textbf {\bibinfo {volume} {1}},\ \bibinfo {pages} {119--121} (\bibinfo {year} {2013})}\BibitemShut {NoStop}%
\bibitem [{\citenamefont {Wilson}, \citenamefont {Stevens},\ and\ \citenamefont {Woodall}(2019)}]{wilson2019modeling}%
  \BibitemOpen
  \bibfield  {author} {\bibinfo {author} {\bibfnamefont {J.~D.}\ \bibnamefont {Wilson}}, \bibinfo {author} {\bibfnamefont {N.~T.}\ \bibnamefont {Stevens}}, \ and\ \bibinfo {author} {\bibfnamefont {W.~H.}\ \bibnamefont {Woodall}},\ }\bibfield  {title} {\enquote {\bibinfo {title} {Modeling and detecting change in temporal networks via the degree corrected stochastic block model},}\ }\href@noop {} {\bibfield  {journal} {\bibinfo  {journal} {Quality and Reliability Engineering International}\ }\textbf {\bibinfo {volume} {35}},\ \bibinfo {pages} {1363--1378} (\bibinfo {year} {2019})}\BibitemShut {NoStop}%
\bibitem [{\citenamefont {Lewis}\ \emph {et~al.}(2019)\citenamefont {Lewis}, \citenamefont {Keith}, \citenamefont {Howard}, \citenamefont {Adam}, \citenamefont {Aaron},\ and\ \citenamefont {Luke}}]{voteview}%
  \BibitemOpen
  \bibfield  {author} {\bibinfo {author} {\bibfnamefont {J.~B.}\ \bibnamefont {Lewis}}, \bibinfo {author} {\bibfnamefont {P.}~\bibnamefont {Keith}}, \bibinfo {author} {\bibfnamefont {R.}~\bibnamefont {Howard}}, \bibinfo {author} {\bibfnamefont {B.}~\bibnamefont {Adam}}, \bibinfo {author} {\bibfnamefont {R.}~\bibnamefont {Aaron}}, \ and\ \bibinfo {author} {\bibfnamefont {S.}~\bibnamefont {Luke}},\ }\href@noop {} {\enquote {\bibinfo {title} {Voteview: Congressional roll-call votes database},}\ }\bibinfo {howpublished} {\url{https://voteview.com/}} (\bibinfo {year} {2019})\BibitemShut {NoStop}%
\bibitem [{\citenamefont {Roy}, \citenamefont {Atchad{\'e}},\ and\ \citenamefont {Michailidis}(2017)}]{roy2017change}%
  \BibitemOpen
  \bibfield  {author} {\bibinfo {author} {\bibfnamefont {S.}~\bibnamefont {Roy}}, \bibinfo {author} {\bibfnamefont {Y.}~\bibnamefont {Atchad{\'e}}}, \ and\ \bibinfo {author} {\bibfnamefont {G.}~\bibnamefont {Michailidis}},\ }\bibfield  {title} {\enquote {\bibinfo {title} {Change point estimation in high dimensional markov random-field models},}\ }\href@noop {} {\bibfield  {journal} {\bibinfo  {journal} {Journal of the Royal Statistical Society: Series B (Statistical Methodology)}\ }\textbf {\bibinfo {volume} {79}},\ \bibinfo {pages} {1187--1206} (\bibinfo {year} {2017})}\BibitemShut {NoStop}%
\bibitem [{Note8()}]{Note8}%
  \BibitemOpen
  \bibinfo {note} {We extensively tested via simulation that, for the model at hand and $T$ and $N$ taken from the data, estimating the variance of the latent parameters in such a way results, on average, in a small underestimation. In checking the coverage of the confidence bands, we considered a DGP with variance increased, with respect to the one estimated on the filtered time series, to compensate for this bias.}\BibitemShut {Stop}%
\bibitem [{\citenamefont {Karahano{\u{g}}lu}\ and\ \citenamefont {Van De~Ville}(2017)}]{karahanouglu2017dynamics}%
  \BibitemOpen
  \bibfield  {author} {\bibinfo {author} {\bibfnamefont {F.~I.}\ \bibnamefont {Karahano{\u{g}}lu}}\ and\ \bibinfo {author} {\bibfnamefont {D.}~\bibnamefont {Van De~Ville}},\ }\bibfield  {title} {\enquote {\bibinfo {title} {Dynamics of large-scale fmri networks: Deconstruct brain activity to build better models of brain function},}\ }\href@noop {} {\bibfield  {journal} {\bibinfo  {journal} {Current Opinion in Biomedical Engineering}\ }\textbf {\bibinfo {volume} {3}},\ \bibinfo {pages} {28--36} (\bibinfo {year} {2017})}\BibitemShut {NoStop}%
\bibitem [{\citenamefont {Simpson}, \citenamefont {Hayasaka},\ and\ \citenamefont {Laurienti}(2011)}]{simpson2011exponential}%
  \BibitemOpen
  \bibfield  {author} {\bibinfo {author} {\bibfnamefont {S.~L.}\ \bibnamefont {Simpson}}, \bibinfo {author} {\bibfnamefont {S.}~\bibnamefont {Hayasaka}}, \ and\ \bibinfo {author} {\bibfnamefont {P.~J.}\ \bibnamefont {Laurienti}},\ }\bibfield  {title} {\enquote {\bibinfo {title} {Exponential random graph modeling for complex brain networks},}\ }\href@noop {} {\bibfield  {journal} {\bibinfo  {journal} {PloS one}\ }\textbf {\bibinfo {volume} {6}},\ \bibinfo {pages} {e20039} (\bibinfo {year} {2011})}\BibitemShut {NoStop}%
\bibitem [{\citenamefont {Giraitis}\ \emph {et~al.}(2016)\citenamefont {Giraitis}, \citenamefont {Kapetanios}, \citenamefont {Wetherilt},\ and\ \citenamefont {{\v{Z}}ike{\v{s}}}}]{giraitis2016estimating}%
  \BibitemOpen
  \bibfield  {author} {\bibinfo {author} {\bibfnamefont {L.}~\bibnamefont {Giraitis}}, \bibinfo {author} {\bibfnamefont {G.}~\bibnamefont {Kapetanios}}, \bibinfo {author} {\bibfnamefont {A.}~\bibnamefont {Wetherilt}}, \ and\ \bibinfo {author} {\bibfnamefont {F.}~\bibnamefont {{\v{Z}}ike{\v{s}}}},\ }\bibfield  {title} {\enquote {\bibinfo {title} {Estimating the dynamics and persistence of financial networks, with an application to the sterling money market},}\ }\href@noop {} {\bibfield  {journal} {\bibinfo  {journal} {Journal of Applied Econometrics}\ }\textbf {\bibinfo {volume} {31}},\ \bibinfo {pages} {58--84} (\bibinfo {year} {2016})}\BibitemShut {NoStop}%
\bibitem [{\citenamefont {Erd{\H{o}}s}\ and\ \citenamefont {R{\'e}nyi}(1959)}]{erdds1959random}%
  \BibitemOpen
  \bibfield  {author} {\bibinfo {author} {\bibfnamefont {P.}~\bibnamefont {Erd{\H{o}}s}}\ and\ \bibinfo {author} {\bibfnamefont {A.}~\bibnamefont {R{\'e}nyi}},\ }\bibfield  {title} {\enquote {\bibinfo {title} {On random graphs i.}}\ }\href@noop {} {\bibfield  {journal} {\bibinfo  {journal} {Publ. Math. Debrecen}\ }\textbf {\bibinfo {volume} {6}},\ \bibinfo {pages} {290--297} (\bibinfo {year} {1959})}\BibitemShut {NoStop}%
\bibitem [{Note9()}]{Note9}%
  \BibitemOpen
  \bibinfo {note} {In the whole paper, we do not allow for links that start and end at the same node, so named \protect \textit {self-loops}. However, including them would be trivial.}\BibitemShut {Stop}%
\bibitem [{\citenamefont {Fienberg}\ and\ \citenamefont {Wasserman}(1981)}]{fienberg1981categorical}%
  \BibitemOpen
  \bibfield  {author} {\bibinfo {author} {\bibfnamefont {S.~E.}\ \bibnamefont {Fienberg}}\ and\ \bibinfo {author} {\bibfnamefont {S.~S.}\ \bibnamefont {Wasserman}},\ }\bibfield  {title} {\enquote {\bibinfo {title} {Categorical data analysis of single sociometric relations},}\ }\href@noop {} {\bibfield  {journal} {\bibinfo  {journal} {Sociological Methodology}\ }\textbf {\bibinfo {volume} {12}},\ \bibinfo {pages} {156--192} (\bibinfo {year} {1981})}\BibitemShut {NoStop}%
\bibitem [{\citenamefont {Wasserman}\ and\ \citenamefont {Pattison}(1996)}]{wasserman1996logit}%
  \BibitemOpen
  \bibfield  {author} {\bibinfo {author} {\bibfnamefont {S.}~\bibnamefont {Wasserman}}\ and\ \bibinfo {author} {\bibfnamefont {P.}~\bibnamefont {Pattison}},\ }\bibfield  {title} {\enquote {\bibinfo {title} {Logit models and logistic regressions for social networks: I. {A}n introduction to markov graphs and p},}\ }\href@noop {} {\bibfield  {journal} {\bibinfo  {journal} {Psychometrika}\ }\textbf {\bibinfo {volume} {61}},\ \bibinfo {pages} {401--425} (\bibinfo {year} {1996})}\BibitemShut {NoStop}%
\bibitem [{\citenamefont {Frank}\ and\ \citenamefont {Strauss}(1986)}]{frank1986markov}%
  \BibitemOpen
  \bibfield  {author} {\bibinfo {author} {\bibfnamefont {O.}~\bibnamefont {Frank}}\ and\ \bibinfo {author} {\bibfnamefont {D.}~\bibnamefont {Strauss}},\ }\bibfield  {title} {\enquote {\bibinfo {title} {Markov graphs},}\ }\href@noop {} {\bibfield  {journal} {\bibinfo  {journal} {Journal of the American Statistical Association}\ }\textbf {\bibinfo {volume} {81}},\ \bibinfo {pages} {832--842} (\bibinfo {year} {1986})}\BibitemShut {NoStop}%
\bibitem [{\citenamefont {Yan}\ \emph {et~al.}(2016)\citenamefont {Yan}, \citenamefont {Leng}, \citenamefont {Zhu} \emph {et~al.}}]{yan2016asymptotics}%
  \BibitemOpen
  \bibfield  {author} {\bibinfo {author} {\bibfnamefont {T.}~\bibnamefont {Yan}}, \bibinfo {author} {\bibfnamefont {C.}~\bibnamefont {Leng}}, \bibinfo {author} {\bibfnamefont {J.}~\bibnamefont {Zhu}},  \emph {et~al.},\ }\bibfield  {title} {\enquote {\bibinfo {title} {Asymptotics in directed exponential random graph models with an increasing bi-degree sequence},}\ }\href@noop {} {\bibfield  {journal} {\bibinfo  {journal} {The Annals of Statistics}\ }\textbf {\bibinfo {volume} {44}},\ \bibinfo {pages} {31--57} (\bibinfo {year} {2016})}\BibitemShut {NoStop}%
\bibitem [{\citenamefont {Yan}\ \emph {et~al.}(2018)\citenamefont {Yan}, \citenamefont {Jiang}, \citenamefont {Fienberg},\ and\ \citenamefont {Leng}}]{yan2018statistical}%
  \BibitemOpen
  \bibfield  {author} {\bibinfo {author} {\bibfnamefont {T.}~\bibnamefont {Yan}}, \bibinfo {author} {\bibfnamefont {B.}~\bibnamefont {Jiang}}, \bibinfo {author} {\bibfnamefont {S.~E.}\ \bibnamefont {Fienberg}}, \ and\ \bibinfo {author} {\bibfnamefont {C.}~\bibnamefont {Leng}},\ }\bibfield  {title} {\enquote {\bibinfo {title} {Statistical inference in a directed network model with covariates},}\ }\href@noop {} {\bibfield  {journal} {\bibinfo  {journal} {Journal of the American Statistical Association}\ ,\ \bibinfo {pages} {1--12}} (\bibinfo {year} {2018})}\BibitemShut {NoStop}%
\bibitem [{\citenamefont {Jochmans}(2018)}]{jochmans2018semiparametric}%
  \BibitemOpen
  \bibfield  {author} {\bibinfo {author} {\bibfnamefont {K.}~\bibnamefont {Jochmans}},\ }\bibfield  {title} {\enquote {\bibinfo {title} {Semiparametric analysis of network formation},}\ }\href@noop {} {\bibfield  {journal} {\bibinfo  {journal} {Journal of Business \& Economic Statistics}\ }\textbf {\bibinfo {volume} {36}},\ \bibinfo {pages} {705--713} (\bibinfo {year} {2018})}\BibitemShut {NoStop}%
\bibitem [{Note10()}]{Note10}%
  \BibitemOpen
  \bibinfo {note} {For a network with $N$ nodes, the number of possible links is of order $N^2$. Instead, when all nodes have a fixed average degree $d$, the number of present links is $d N$, and the density is of order $1/N$.}\BibitemShut {Stop}%
\bibitem [{Note11()}]{Note11}%
  \BibitemOpen
  \bibinfo {note} {Examples of such statistics are the count of 2 stars present in the network or the number of triangles ~\protect \citep {wasserman1996logit}.}\BibitemShut {Stop}%
\bibitem [{\citenamefont {Handcock}\ \emph {et~al.}(2003)\citenamefont {Handcock}, \citenamefont {Robins}, \citenamefont {Snijders}, \citenamefont {Moody},\ and\ \citenamefont {Besag}}]{handcock2003assessing}%
  \BibitemOpen
  \bibfield  {author} {\bibinfo {author} {\bibfnamefont {M.~S.}\ \bibnamefont {Handcock}}, \bibinfo {author} {\bibfnamefont {G.}~\bibnamefont {Robins}}, \bibinfo {author} {\bibfnamefont {T.}~\bibnamefont {Snijders}}, \bibinfo {author} {\bibfnamefont {J.}~\bibnamefont {Moody}}, \ and\ \bibinfo {author} {\bibfnamefont {J.}~\bibnamefont {Besag}},\ }\href@noop {} {\enquote {\bibinfo {title} {Assessing degeneracy in statistical models of social networks},}\ }\bibinfo {type} {Tech. Rep.}\ (\bibinfo  {institution} {Working paper},\ \bibinfo {year} {2003})\BibitemShut {NoStop}%
\bibitem [{\citenamefont {Handcock}(2003)}]{handcock2003statistical}%
  \BibitemOpen
  \bibfield  {author} {\bibinfo {author} {\bibfnamefont {M.~S.}\ \bibnamefont {Handcock}},\ }\bibfield  {title} {\enquote {\bibinfo {title} {Statistical models for social networks: Inference and degeneracy},}\ }in\ \href@noop {} {\emph {\bibinfo {booktitle} {Dynamic Social Network Modeling and Analysis: Workshop Summary and Papers}}}\ (\bibinfo {organization} {National Academies Press},\ \bibinfo {year} {2003})\ p.\ \bibinfo {pages} {229}\BibitemShut {NoStop}%
\bibitem [{\citenamefont {Schweinberger}(2011)}]{schweinberger2011instability}%
  \BibitemOpen
  \bibfield  {author} {\bibinfo {author} {\bibfnamefont {M.}~\bibnamefont {Schweinberger}},\ }\bibfield  {title} {\enquote {\bibinfo {title} {Instability, sensitivity, and degeneracy of discrete exponential families},}\ }\href@noop {} {\bibfield  {journal} {\bibinfo  {journal} {Journal of the American Statistical Association}\ }\textbf {\bibinfo {volume} {106}},\ \bibinfo {pages} {1361--1370} (\bibinfo {year} {2011})}\BibitemShut {NoStop}%
\bibitem [{Note12()}]{Note12}%
  \BibitemOpen
  \bibinfo {note} {\protect \textit {Edgewise} precisely means that we count partners only if shared by nodes that are connected.}\BibitemShut {Stop}%
\bibitem [{\citenamefont {Strauss}\ and\ \citenamefont {Ikeda}(1990)}]{strauss1990pseudolikelihood}%
  \BibitemOpen
  \bibfield  {author} {\bibinfo {author} {\bibfnamefont {D.}~\bibnamefont {Strauss}}\ and\ \bibinfo {author} {\bibfnamefont {M.}~\bibnamefont {Ikeda}},\ }\bibfield  {title} {\enquote {\bibinfo {title} {Pseudolikelihood estimation for social networks},}\ }\href@noop {} {\bibfield  {journal} {\bibinfo  {journal} {Journal of the American Statistical Association}\ }\textbf {\bibinfo {volume} {85}},\ \bibinfo {pages} {204--212} (\bibinfo {year} {1990})}\BibitemShut {NoStop}%
\bibitem [{\citenamefont {Varin}, \citenamefont {Reid},\ and\ \citenamefont {Firth}(2011)}]{varin2011overview}%
  \BibitemOpen
  \bibfield  {author} {\bibinfo {author} {\bibfnamefont {C.}~\bibnamefont {Varin}}, \bibinfo {author} {\bibfnamefont {N.}~\bibnamefont {Reid}}, \ and\ \bibinfo {author} {\bibfnamefont {D.}~\bibnamefont {Firth}},\ }\bibfield  {title} {\enquote {\bibinfo {title} {An overview of composite likelihood methods},}\ }\href@noop {} {\bibfield  {journal} {\bibinfo  {journal} {Statistica Sinica}\ ,\ \bibinfo {pages} {5--42}} (\bibinfo {year} {2011})}\BibitemShut {NoStop}%
\bibitem [{\citenamefont {Van~Duijn}, \citenamefont {Gile},\ and\ \citenamefont {Handcock}(2009)}]{vanduijn2009}%
  \BibitemOpen
  \bibfield  {author} {\bibinfo {author} {\bibfnamefont {M.~A.}\ \bibnamefont {Van~Duijn}}, \bibinfo {author} {\bibfnamefont {K.~J.}\ \bibnamefont {Gile}}, \ and\ \bibinfo {author} {\bibfnamefont {M.~S.}\ \bibnamefont {Handcock}},\ }\bibfield  {title} {\enquote {\bibinfo {title} {A framework for the comparison of maximum pseudo-likelihood and maximum likelihood estimation of exponential family random graph models},}\ }\href@noop {} {\bibfield  {journal} {\bibinfo  {journal} {Social Networks}\ }\textbf {\bibinfo {volume} {31}},\ \bibinfo {pages} {52 -- 62} (\bibinfo {year} {2009})}\BibitemShut {NoStop}%
\bibitem [{\citenamefont {Desmarais}\ and\ \citenamefont {Cranmer}(2012)}]{desmarais2012statistical}%
  \BibitemOpen
  \bibfield  {author} {\bibinfo {author} {\bibfnamefont {B.~A.}\ \bibnamefont {Desmarais}}\ and\ \bibinfo {author} {\bibfnamefont {S.~J.}\ \bibnamefont {Cranmer}},\ }\bibfield  {title} {\enquote {\bibinfo {title} {Statistical mechanics of networks: Estimation and uncertainty},}\ }\href@noop {} {\bibfield  {journal} {\bibinfo  {journal} {Physica A: Statistical Mechanics and its Applications}\ }\textbf {\bibinfo {volume} {391}},\ \bibinfo {pages} {1865--1876} (\bibinfo {year} {2012})}\BibitemShut {NoStop}%
\bibitem [{\citenamefont {Schmid}\ and\ \citenamefont {Desmarais}(2017)}]{schmid2017exponential}%
  \BibitemOpen
  \bibfield  {author} {\bibinfo {author} {\bibfnamefont {C.~S.}\ \bibnamefont {Schmid}}\ and\ \bibinfo {author} {\bibfnamefont {B.~A.}\ \bibnamefont {Desmarais}},\ }\bibfield  {title} {\enquote {\bibinfo {title} {Exponential random graph models with big networks: Maximum pseudolikelihood estimation and the parametric bootstrap},}\ }in\ \href@noop {} {\emph {\bibinfo {booktitle} {Big Data (Big Data), 2017 IEEE International Conference on}}}\ (\bibinfo {organization} {IEEE},\ \bibinfo {year} {2017})\ pp.\ \bibinfo {pages} {116--121}\BibitemShut {NoStop}%
\bibitem [{\citenamefont {Blasques}, \citenamefont {Koopman},\ and\ \citenamefont {Lucas}(2014)}]{blasques2014maximum}%
  \BibitemOpen
  \bibfield  {author} {\bibinfo {author} {\bibfnamefont {F.}~\bibnamefont {Blasques}}, \bibinfo {author} {\bibfnamefont {S.~J.}\ \bibnamefont {Koopman}}, \ and\ \bibinfo {author} {\bibfnamefont {A.}~\bibnamefont {Lucas}},\ }\href@noop {} {\enquote {\bibinfo {title} {Maximum likelihood estimation for generalized autoregressive score models},}\ }\bibinfo {type} {Tech. Rep.}\ (\bibinfo  {institution} {Tinbergen Institute Discussion Paper},\ \bibinfo {year} {2014})\BibitemShut {NoStop}%
\bibitem [{\citenamefont {Kullback}(1997)}]{kullback1997information}%
  \BibitemOpen
  \bibfield  {author} {\bibinfo {author} {\bibfnamefont {S.}~\bibnamefont {Kullback}},\ }\href@noop {} {\emph {\bibinfo {title} {Information theory and statistics}}}\ (\bibinfo  {publisher} {Courier Corporation},\ \bibinfo {year} {1997})\BibitemShut {NoStop}%
\bibitem [{\citenamefont {Blasques}, \citenamefont {Lucas},\ and\ \citenamefont {van Vlodrop}(2020)}]{blasques2020finite}%
  \BibitemOpen
  \bibfield  {author} {\bibinfo {author} {\bibfnamefont {F.}~\bibnamefont {Blasques}}, \bibinfo {author} {\bibfnamefont {A.}~\bibnamefont {Lucas}}, \ and\ \bibinfo {author} {\bibfnamefont {A.~C.}\ \bibnamefont {van Vlodrop}},\ }\bibfield  {title} {\enquote {\bibinfo {title} {Finite sample optimality of score-driven volatility models: Some {M}onte {C}arlo evidence},}\ }\href@noop {} {\bibfield  {journal} {\bibinfo  {journal} {Econometrics and Statistics}\ } (\bibinfo {year} {2020})}\BibitemShut {NoStop}%
\bibitem [{\citenamefont {Blasques}\ \emph {et~al.}(2016)\citenamefont {Blasques}, \citenamefont {Koopman}, \citenamefont {{\L}asak},\ and\ \citenamefont {Lucas}}]{blasques2016confbands}%
  \BibitemOpen
  \bibfield  {author} {\bibinfo {author} {\bibfnamefont {F.}~\bibnamefont {Blasques}}, \bibinfo {author} {\bibfnamefont {S.~J.}\ \bibnamefont {Koopman}}, \bibinfo {author} {\bibfnamefont {K.}~\bibnamefont {{\L}asak}}, \ and\ \bibinfo {author} {\bibfnamefont {A.}~\bibnamefont {Lucas}},\ }\bibfield  {title} {\enquote {\bibinfo {title} {In-sample confidence bands and out-of-sample forecast bands for tvps in observation-driven models},}\ }\href@noop {} {\bibfield  {journal} {\bibinfo  {journal} {International Journal of Forecasting}\ }\textbf {\bibinfo {volume} {32}},\ \bibinfo {pages} {875--887} (\bibinfo {year} {2016})}\BibitemShut {NoStop}%
\bibitem [{\citenamefont {Huber}\ \emph {et~al.}(1967)\citenamefont {Huber} \emph {et~al.}}]{huber1967behavior}%
  \BibitemOpen
  \bibfield  {author} {\bibinfo {author} {\bibfnamefont {P.~J.}\ \bibnamefont {Huber}} \emph {et~al.},\ }\bibfield  {title} {\enquote {\bibinfo {title} {The behavior of maximum likelihood estimates under nonstandard conditions},}\ }in\ \href@noop {} {\emph {\bibinfo {booktitle} {Proceedings of the fifth Berkeley symposium on mathematical statistics and probability}}},\ Vol.~\bibinfo {volume} {1}\ (\bibinfo {organization} {University of California Press},\ \bibinfo {year} {1967})\ pp.\ \bibinfo {pages} {221--233}\BibitemShut {NoStop}%
\bibitem [{\citenamefont {White}(1980)}]{white1980heteroskedasticity}%
  \BibitemOpen
  \bibfield  {author} {\bibinfo {author} {\bibfnamefont {H.}~\bibnamefont {White}},\ }\bibfield  {title} {\enquote {\bibinfo {title} {A heteroskedasticity-consistent covariance matrix estimator and a direct test for heteroskedasticity},}\ }\href@noop {} {\bibfield  {journal} {\bibinfo  {journal} {Econometrica: Journal of the Econometric Society}\ ,\ \bibinfo {pages} {817--838}} (\bibinfo {year} {1980})}\BibitemShut {NoStop}%
\bibitem [{\citenamefont {Hamilton}(1986)}]{HAMILTON1986387}%
  \BibitemOpen
  \bibfield  {author} {\bibinfo {author} {\bibfnamefont {J.~D.}\ \bibnamefont {Hamilton}},\ }\bibfield  {title} {\enquote {\bibinfo {title} {A standard error for the estimated state vector of a state-space model},}\ }\href {\doibase https://doi.org/10.1016/0304-4076(86)90004-7} {\bibfield  {journal} {\bibinfo  {journal} {Journal of Econometrics}\ }\textbf {\bibinfo {volume} {33}},\ \bibinfo {pages} {387--397} (\bibinfo {year} {1986})}\BibitemShut {NoStop}%
\bibitem [{\citenamefont {Davidson}, \citenamefont {MacKinnon}\ \emph {et~al.}(2004)\citenamefont {Davidson}, \citenamefont {MacKinnon} \emph {et~al.}}]{davidson2004econometric}%
  \BibitemOpen
  \bibfield  {author} {\bibinfo {author} {\bibfnamefont {R.}~\bibnamefont {Davidson}}, \bibinfo {author} {\bibfnamefont {J.~G.}\ \bibnamefont {MacKinnon}},  \emph {et~al.},\ }\href@noop {} {\emph {\bibinfo {title} {Econometric theory and methods}}}\ (\bibinfo  {publisher} {Oxford University Press New York},\ \bibinfo {year} {2004})\BibitemShut {NoStop}%
\bibitem [{\citenamefont {Lee}(1991)}]{lee1991lagrange}%
  \BibitemOpen
  \bibfield  {author} {\bibinfo {author} {\bibfnamefont {J.~H.}\ \bibnamefont {Lee}},\ }\bibfield  {title} {\enquote {\bibinfo {title} {A {L}agrange multiplier test for {GARCH} models},}\ }\href@noop {} {\bibfield  {journal} {\bibinfo  {journal} {Economics Letters}\ }\textbf {\bibinfo {volume} {37}},\ \bibinfo {pages} {265--271} (\bibinfo {year} {1991})}\BibitemShut {NoStop}%
\end{thebibliography}

%
%

\end{document}